\documentclass{aa}  

\usepackage[english]{babel}
\usepackage{graphicx,amsmath,amssymb,gensymb}
\usepackage{txfonts}

\usepackage{hyperref}

\usepackage[singlelinecheck=true,format=default,font=small,labelfont=bf]{caption}
\usepackage{color}

\usepackage{subcaption}

\usepackage[switch, modulo]{lineno}

\begin{document}

\title{The Fornax 3D project: dust mix and gas properties in the center of early-type galaxy FCC~167}

\author{S.~Viaene\inst{\ref{inst-UHerts}, \ref{inst-UGent}, \thanks{\email{sebastien.viaene@ugent.be}}}
\and M.~Sarzi\inst{\ref{inst-Armagh},\ref{inst-UHerts}}
\and N.~Zabel\inst{\ref{inst-Cardiff}}
\and L.~Coccato\inst{\ref{inst-ESO}}
\and E.~M.~Corsini\inst{\ref{inst-Padova},\ref{inst-INAF}}
\and T.~A.~Davis\inst{\ref{inst-Cardiff}}
\and P.~De Vis\inst{\ref{inst-IAS}}
\and P.~T.~de~Zeeuw\inst{\ref{inst-Leiden},\ref{inst-MPIE}}
\and J.~Falcón-Barroso\inst{\ref{inst-IAC},\ref{inst-Tenerife}}
\and D.A.~Gadotti\inst{\ref{inst-ESO}}
\and E.~Iodice\inst{\ref{inst-Napoli}}
\and M.~Lyubenova\inst{\ref{inst-ESO}}
\and R.~McDermid\inst{\ref{inst-Sydney},\ref{inst-AAO}}
\and L.~Morelli\inst{\ref{inst-Chile},\ref{inst-Padova},\ref{inst-INAF}}
\and B.~Nedelchev\inst{\ref{inst-UHerts}}
\and F.~Pinna\inst{\ref{inst-IAC},\ref{inst-Tenerife}}
\and T.~W.~Spriggs\inst{\ref{inst-UHerts}}
\and G.~van~de~Ven\inst{\ref{inst-ESO}}
}
\institute{Centre for Astrophysics Research, University of Hertfordshire, College Lane, Hatfield AL10 9AB, UK  \label{inst-UHerts}
\and Sterrenkundig Observatorium, Universiteit Gent, Krijgslaan 281, B-9000 Gent, Belgium \label{inst-UGent}
\and Armagh Observatory and Planetarium, College Hill, Armagh, BT61 9DG, UK  \label{inst-Armagh}
\and School of Physics \& Astronomy, Cardiff University, Queens Buildings, The Parade, Cardiff, CF24 3AA, UK \label{inst-Cardiff} 
\and European Southern Observatory, Karl-Schwarzschild-Str. 2, 85748 Garching b. München, Germany  \label{inst-ESO}
\and Dipartimento di Fisica e Astronomia `G. Galilei', Università di Padova, vicolo dell'Osservatorio 3, I-35122 Padova, Italy \label{inst-Padova}
\and INAF–Osservatorio Astronomico di Padova, vicolo dell'Osservatorio 5, I-35122 Padova, Italy \label{inst-INAF}
\and Institut d'Astrophysique Spatiale, CNRS, Université Paris-Sud, Université Paris-Saclay, Bat. 121, F-91405 Orsay Cedex, France \label{inst-IAS} 
\and Sterrewacht Leiden, Leiden University, Postbus 9513, 2300 RA Leiden, The Netherlands \label{inst-Leiden}
\and Max-Planck-Institut fuer extraterrestrische Physik, Giessenbachstrasse, 85741 Garching bei Muenchen, Germany \label{inst-MPIE}
\and Instituto de Astrofísica de Canarias, C/ Via Láctea s/n, E-38200 La Laguna, Tenerife, Spain \label{inst-IAC}
\and Depto. Astrofísica, Universidad de La Laguna (ULL), E-38206 La Laguna, Tenerife, Spain \label{inst-Tenerife}
\and INAF–Osservatorio Astronomico di Capodimonte, via Moiariello 16, I-80131 Napoli, Italy  \label{inst-Napoli}
\and Department of Physics and Astronomy, Macquarie University, Sydney, NSW 2109, Australia \label{inst-Sydney}
\and Australian Astronomical Observatory, PO Box 915, Sydney, NSW 1670, Australia \label{inst-AAO}
\and Instituto de Astronomía y Ciencias Planetarias, Universidad de Atacama, Copayapu 485, Copiapó , Chile  \label{inst-Chile}
}

\abstract{
Galaxies continuously reprocess their interstellar material. One can therefore expect changing dust grain properties in galaxies which have followed different evolutionary pathways. Determining the intrinsic dust grain mix of a galaxy helps in reconstructing its evolutionary history. Early-type galaxies occasionally display regular dust lanes in their central regions. Due to the relatively simple geometry and composition of their stellar bodies, these galaxies are ideal to disentangle dust mix variations from geometric effects. We therefore model the various components of such a galaxy (FCC~167). We reconstruct its recent history, and investigate the possible fate of the dust lane. MUSE and ALMA observations reveal a nested ISM structure. An ionised-gas disk pervades the central regions of FCC~167, including those occupied by the main dust lane. Inward of the dust lane, we also find a disk/ring of cold molecular gas where stars are forming and HII regions contribute to the ionised-gas emission. Further in, the gas ionisation points towards an active galactic nucleus and the fuelling of a central supermassive black hole from its surrounding ionised and molecular reservoir. Observational constraints and radiative transfer models suggest the dust and gas are distributed in a ring-like geometry and the dust mix lacks small grains. The derived dust destruction timescales from sputtering in hot gas are short and we conclude that the dust must be strongly self-shielding and clumpy, or will quickly be eroded and disappear. Our findings show how detailed analysis of individual systems can complement statistical studies of dust-lane ETGs.
}


\titlerunning{F3D: ISM in FCC~167}
\authorrunning{S. Viaene}

\maketitle

\section{Introduction}

Galaxies are inherently three-dimensional (3D) and much of their internal structure is hidden in their projection on the sky. 
The presence of dust and gas in galaxies of any type further complicates the measurement of their intrinsic shape, and the properties of their stellar populations such as mass, age, metallicity and initial mass function (IMF). Such information is already blended along the line-of-sight and in addition effectively obscured by dust. Interstellar dust grains both absorb and scatter light from the ultraviolet (UV) to the near-infrared (NIR), with higher efficiency for shorter wavelengths. 

At the same time however, dust is a vital ingredient in galaxy evolution. As dust grains are intermixed with the gas, collisions induce thermal equilibrium between these components \citep{Croxall2012, Hughes2015}. Gas can cool by transferring kinetic energy to the dust grains, which can more easily convert this into radiative energy at far-infrared (FIR) to millimetre wavelengths. In addition, it is on the dust grains that H$_2$ and other molecules form most efficiently, speeding up molecular cloud formation \citep{Hollenbach1971,Vidali2004,Bron2014}. Finally, the already formed molecules are then shielded by the dust as harmful radiation from stars causes photo-dissociation \citep[see e.g.][]{Draine1996,Walch2015}. Taken together, dust thus plays a vital role in the star formation process, and by tracking dust in different environments, one can investigate the efficiency of these processes.

On average, one third of starlight in galaxies is absorbed and re-emitted at far-infrared (FIR) and submillimeter (submm) wavelengths \citep{Popescu2002, Skibba2011, Viaene2016}. For early-type galaxies (ETGs), this is usually less \citep[see e.g.][]{Bianchi2018}. About $4\%$ of ETGs show large-scale dust absorption features in optical SDSS images \citep{Kaviraj2012,Kaviraj2013}. A larger fraction ($22 \%$) do contain molecular gas \citep{Young2011}, which is usually associated with dust absorption features in high-resolution optical imaging \citep[e.g. ][]{Ferrarese2006}. Such dust-lane ETGs occupy the phase-space between actively star-forming galaxies, and passive, massive systems. The minor-merger scenario has long been proposed as the most likely mechanism to explain the presence of dust (and gas) in their interstellar medium (ISM, see e.g. \citealt{Bertola1992}) In this scenario, small, gas-rich galaxies merge with passive ETGs. These events are thought to spark star-formation leading to a complex ISM on short timescales \citep[see e.g.][]{Kaviraj2007, Kaviraj2009, Kaviraj2013, Davis2015}. It is unclear how long the effects of a minor merger remain visible. On the one hand, the presence of dust lanes suggest that the ISM has had the time to settle in disks or rings. On the other hand, dust that is not shielded should not survive long in the harsh radiation field of ETGs, many of which are embedded in X-ray halos \citep[see e.g.][]{Fabbiano1989, Anderson2015} or display central far-UV emission \citep[e.g.][]{Burstein1988}. To address these questions, it is important to study individual objects in detail, in complement with the statistical studies mentioned above.

In this paper, we follow the conventional distinction between extinction and attenuation. The combination of absorption and scattering out of the line of sight is defined as dust extinction and effectively reddens the spectrum of a galaxy. Dust attenuation is then defined as the combination of extinction and scattering back into the line of sight. Attenuation therefore depends on the global distribution of stars and dust in a galaxy. It can flatten the intrinsic reddening law as blue light is more efficiently scattered. A three-dimensional (3D) reconstruction of a galaxy and its intrinsic stellar properties thus requires an adequate correction of dust attenuation.

At the same time, the wavelength-dependent attenuation (the attenuation curve) holds information on the dust mixture itself \citep[see][for a review]{Galliano2018}. A steep attenuation curve can for example indicate an important population of small dust grains \citep{Draine2003, Jones2013}. However, this information is encrypted by geometric effects. For example, it can also indicate that (along the line-of-sight) a higher fraction of the stars lie behind the dust \citep[see e.g.][]{Viaene2017}. It is important to break this degeneracy and learn more about the intrinsic dust mix in each galaxy, which is often assumed to be the same for all galaxies. 

There are three main modes of dust production (winds of asymptotic giant branch -- AGB -- stars, supernova ejecta and ISM grain growth) which can all vary in importance for different galaxy types \citep[see][and references therein]{Tielens2005, Finkelman2012, Rowlands2014, Hirashita2015, DeVis2017, Zhukovska2018}. There is also a plethora of dust destruction processes \citep{Jones2004} depending on the local environment. It is unlikely that dust production and destruction is self-regulating for the whole galaxy population and thus differences in the intrinsic dust mix are expected from galaxy to galaxy, and even inside a single galaxy.

To approach this complex problem, one first requires solid measurements of the attenuation levels inside a galaxy, and construct an absolute attenuation curve. This has been difficult in the past, as classic extinction studies \citep[applied in the Milky Way and Magellanic clouds by e.g. ][]{Cardelli1989, Gordon2003,Fitzpatrick2007} are not possible when individual stars cannot be resolved. As such, measurements of the attenuation curve of galaxies is often done in a statistical way, based on broad-band colours, most notably the UV slope, or through the Balmer decrement \citep[see e.g.][and references therein]{Calzetti2000, Wild2011, Battisti2016}. Only the dust-lane ETGs allow an easy empirical measurement of their attenuation profile due to their smooth stellar body and strong, localised absorption features. Historically, their attenuation levels can be determined by constructing a dust-free model from broad-band images and comparing that to the observed, reddened galaxy \citep{Goudfrooij1994b, Patil2007, Finkelman2008, Finkelman2010, Kaviraj2012, Kulkarni2014, Viaene2015}. 

In addition, state-of-the-art spectral fitting can provide an independent measure of the attenuation levels when properly modelled. This requires a two-component reddening law, which differentiates between line attenuation (by dust in star-forming regions) and continuum attenuation (by diffuse dust). This method was first introduced for broad-band modeling of spectral energy distributions (SEDs, \citealt{Silva1998, Charlot2000}) and only later introduced in spectral modeling \citep[see e.g.][]{Oh2011}. However, the useful comparison of spectral-fitting attenuation with empirical measurements (as outlined above) has not been made to our knowledge

\begin{figure*}
	\includegraphics[width=\textwidth]{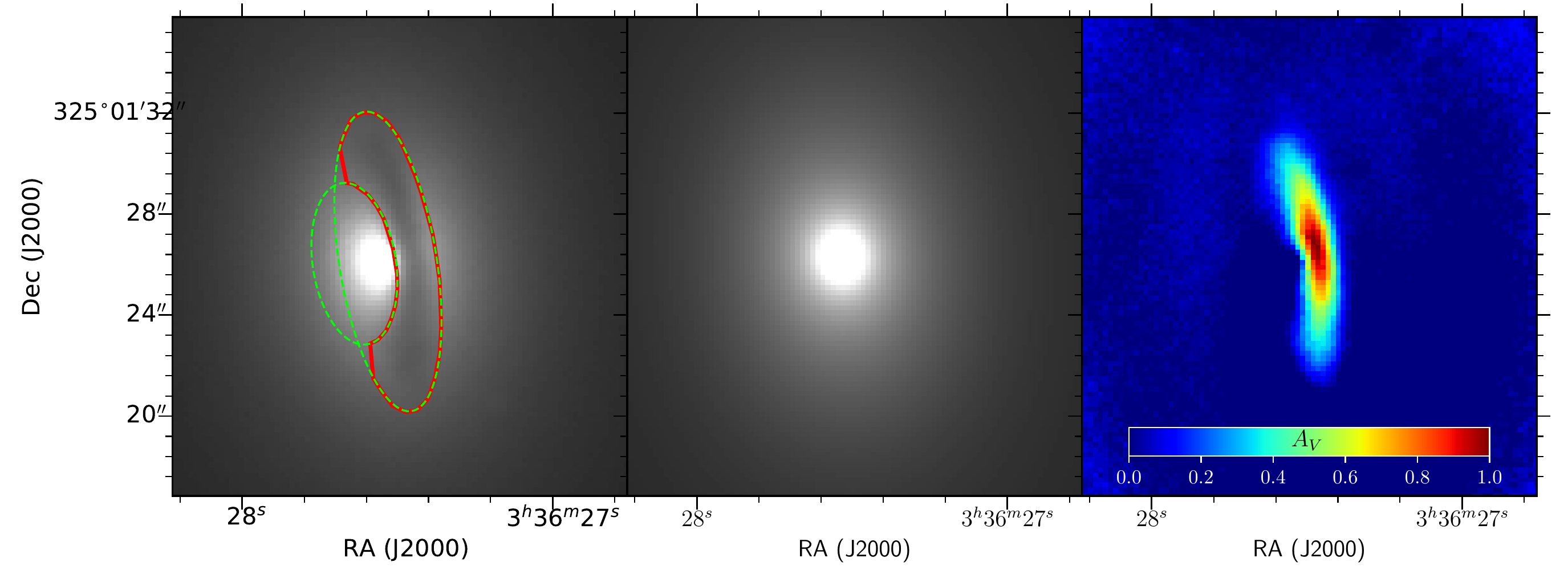}
    \caption{Different stages of the MGE modeling to produce attenuation maps for every wavelength slice of the MUSE datacube. Left: reconstructed image at $5509 \, \AA$ (the $V$-band central wavelength). The green ellipses mark the inner and outer edge of the dust lane (red contour). Middle: the corresponding dust-free MGE model fit. Right: the attenuation map at $5509\, \AA$ obtained from Eq.~\ref{eq:att}.}
    \label{fig:MGEfit}
\end{figure*}

Most recently, we applied this empirical method to NGC~5626, a dust-lane ETG \citep{Viaene2017}. Instead of broad-band imaging, however, we made use of integral-field spectroscopic data from MUSE \citep{MUSE}. This essentially provides a high-quality image of the galaxy every few $\text{\AA}$ngstrom, spanning a wide range in optical wavelength. As such, the first absolute attenuation curve at high spectral resolution could be measured for an external galaxy. In this paper, we complement this study by targeting a dust-lane ETG in the Fornax Cluster: FCC~167 (NGC 1380). It is one of the few dust-lane ETGs with MUSE data available, and contains only a circum-nuclear ISM, making it more comparable to the majority of the gas-rich ETG population \citep{Young2011, Davis2013}. The dust in FCC~167 could therefore be quite different from the large-scale dust lane observed in NGC~5626. 

The additional advantage of IFU observations is that the morphology and kinematics of the ionised gas can be studied simultaneously. The total flux of the ionised gas highlights its projected morphology, while the kinematic moment maps reveal its dynamics. This offers an insight in the 3D distribution of the ISM. The ionised gas may be diffuse or located in photodissociation regions (PDRs) around new stars. In both cases the gas can be associated with dust and thus set a prior on the geometry of the dust component. This can in turn be used to select a limited set of possible 3D dust distributions and feed them into radiative transfer simulations. Only by solving the radiative transfer equations in 3D is it possible to take non-local absorption and scattering (in and out of the line-of-sight) into account. In this way, dust properties and dust distribution can be disentangled from each other. This was successfully applied by \citet{Viaene2015} to NGC 4370, a dust-lane ETG in the Virgo Cluster. They found that a large-scale ring of dust was able to explain the optical attenuation, and was consistent with dust mass estimates from far-infrared (FIR) emission. 

The goal of this paper is to explore whether it is possible to infer qualitative information about the kind of dust in a galaxy by combining constraints on the geometry of the ISM and advanced radiative transfer modeling. This in turn is informed by an accurate and spatially resolved measurement for the attenuation curve and by the ionised-gas kinematics. While answering this question, we also hope to better understand the origin of the ISM within FCC~167. The observations and data analysed here are discussed in Sect.~\ref{sec:data}. In Sect.~\ref{sec:attenuation} we outline our measurements of the attenuation in the dust lane and we discuss other ISM tracers in \ref{sec:ISM}. Radiative transfer modeling of the central regions of FCC~167 is presented in Sect.~\ref{sec:RTmodel} before a more general discussion in Sect.~\ref{sec:discussion}. Finally, we formulate our conclusions in Sect.~\ref{sec:conclusions}.

\section{Observations and Data} \label{sec:data}

FCC~167 ($\alpha_{J2000}=03^h36^m27.55^s$, $\delta_{J2000}=-34\degree58^{\prime}33.88^{\prime\prime}$, $D=18.8$ Mpc) is part of the Fornax Cluster and was observed in the context of the Fornax 3D project (F3D, PI: M. Sarzi). This is a Director's Discretionary Time program on the Multi Unit Spectroscopic Explorer (MUSE, \citealt{MUSE}), an integral-field spectrograph mounted on the Unit Telescope 4 of the Very Large Telescope (VLT). It is a magnitude-limited survey ($m_r > 17$) of all galaxies within the virial radius of the Fornax Cluster. The central pointing of FCC~167, used in this paper, was observed on Dec 31st 2016 in five separate exposures of 720 seconds. Each subsequent exposure was rotated by 90$\degree$ to mitigate instrumental effects. The data reduction was performed using the MUSE pipeline (version 1.6.2, \citealt{Weilbacher2016}). Additional cleaning of contaminating sky lines was performed using the Zurich Atmospheric Purge (ZAP, \citealt{Soto2016}) algorithm. For more information on the observing strategy and data reduction process, we refer to the F3D survey paper \citep{Sarzi2018}. We measure a full width half maximum (FWHM) of $4$ MUSE pixels or $0.8$ arcsec for the instrument point spread function (PSF), corresponding to a physical resolution of $70$ pc.

During this investigation, we also made use of several ancillary observations of FCC~167 (see Sect.~\ref{sec:ISM} and \ref{sec:RTmodel}). To estimate the structure and content of molecular gas we rely on ALMA observations from the ALMA Fornax Cluster Survey (PI: T. Davis). They map the CO(1-0) transition in the galaxy. For details of the survey including moment maps we refer to \citet{Zabel2018}. To constrain the dust heating by hot gas electrons and the associated dust survival timescale we investigate archival X-ray data from the \textit{Chandra} X-ray Observatory.

Finally, our panchromatic radiative transfer models predict the emission in the FIR and submm wavelength regime. To estimate the energy balance in the galaxy and the dust heating mechanisms, we compare the model SED to fluxes from the \textit{Herschel} space telescope \citep{Herschel}. These fluxes were measured from the PACS \citep{PACS} and SPIRE \citep{SPIRE} images provided by the DustPedia \citep{Davies2017} database. We performed our own aperture photometry  using the Python \texttt{photutils} package \citep{photutils} to match the MUSE field-of-view and adequately compare the model and observed fluxes.

\section{Spectrally resolved optical attenuation} \label{sec:attenuation}

\begin{figure*}
	\includegraphics[width=\textwidth]{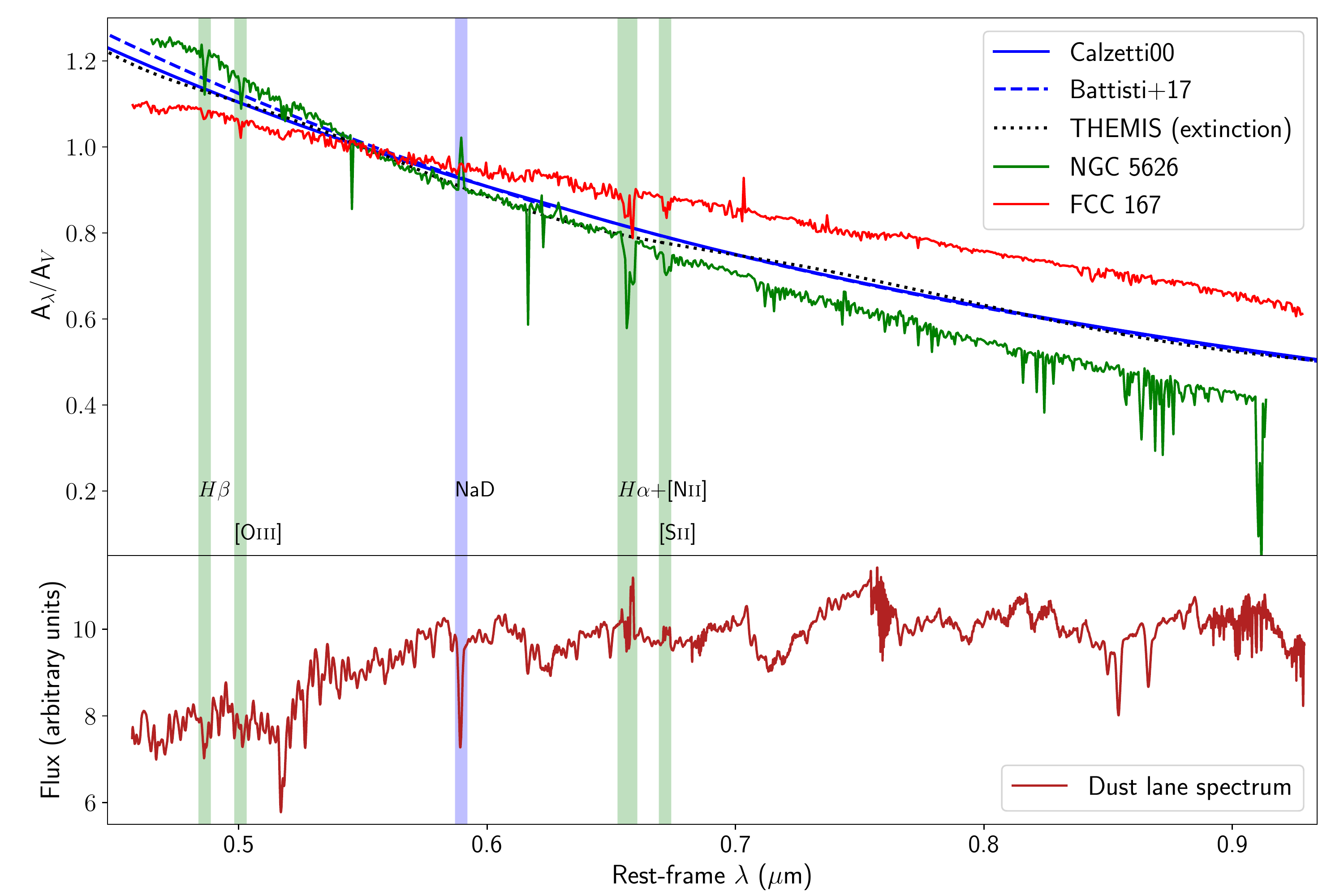}
    \caption{Spectrally resolved attenuation curve for FCC~167 (red line, this work) and for NGC~5626 (green line, \citealt{Viaene2017}). Three reference curves are plotted: The \citet{Calzetti2000} and \citet{Battisti2017} average attenuation curves for different samples of galaxies (blue solid and dashed line, respectively), and the THEMIS \textit{extinction} curve (black dotted line). All curves are normalised to the $V$-band central wavelength, with $A_V = 0.35$ for FCC~167. For reference, the emission spectrum of the dust lane area is show in the bottom panel. Several important emission lines (shaded green) and the NaD absorption line (shaded purple) are highlighted as well.}
    \label{fig:globalAtt}
\end{figure*}

The attenuation along a line-of-sight is defined as the logarithmic difference of observed ($I_\lambda^\text{obs}$) and intrinsic ($I_\lambda^\text{intr}$) intensity of light: 
\begin{equation} \label{eq:att}
A_\lambda = -2.5\log(I_\lambda^\text{obs}/I_\lambda^\text{intr}). 
\end{equation}
Measuring this value is particularly difficult because the intrinsic flux is not known. However, it is possible to estimate this value by interpolating the surface brightness of dust-free regions in the galaxy, or by finding a model for the unattenuated spectrum. The former case requires a smooth stellar body, which is why this method is primarily used for ETGs \citep{Goudfrooij1994b, Patil2007, Finkelman2008, Finkelman2010, Kaviraj2012, Kulkarni2014, Viaene2015}. The latter is more broadly applicable, but works best when spectra (rather than broad-band fluxes) are fitted \citep{Sarzi2006}.

A zoom of the central pointing of FCC~167 is presented in Fig.~\ref{fig:MGEfit} (left panel) and shows the symmetric and smooth stellar surface brightness distribution. The dark dust lane disturbs this distribution but is in itself also symmetric and regular, making this an ideal test case to study circumnuclear dust. We combine two ellipses as projected inner and outer boundaries of the dust lane to construct a mask. The dust lane is not visible on the other side of the major axis due to its high inclination. We do not mask these corresponding pixels as the expected attenuation on the stellar flux is minimal. We use the remaining, unmasked pixels to construct a dust-free model of the galaxy. The same technique as in \citet{Viaene2017} is used, which we briefly summarise here.

The MUSE datacube is binned by a factor of five in wavelength space to boost signal-to-noise ratio, creating independent slices of $\Delta\lambda = 6.25\, \AA$. Each slice is then treated as a separate image, with the dust lane masked (Fig.~\ref{fig:MGEfit}, left panel). A Multi-Gaussian Expansion (MGE, \citealt{Emsellem1994, Cappellari2002}) fit is then performed to create a dust-free image from the unmasked pixels (Fig~\ref{fig:MGEfit} middle panel). The global attenuation in this wavelength bin is then computed by measuring the flux in the dust lane area for both the observed and dust-free image, and applying Eq.~\ref{eq:att}. At the same time, attenuation maps are created for every wavelength bin by subtracting both images in log space (following Eq.~\ref{eq:att} per pixel). An example attenuation map is shown in the right panel of Fig.~\ref{fig:MGEfit}, highlighting the enhanced attenuation in the dust lane area.

The combined global attenuation measurements (in the dust-lane area) are shown in Fig.~\ref{fig:globalAtt} (top panel, red line). For reference, we also plot the total flux in the dust lane in the bottom panel of Fig.~\ref{fig:globalAtt}. Although our method provides an absolute measure of the attenuation in every wavelength bin, we normalise this by the $5509 \, \AA$ bin, to replicate the often shown $A_\lambda/A_V$ ratio. This allows us to compare the observed attenuation curve in FCC~167 with literature curves. The most direct comparison to make is the only other spectrally resolved attenuation curve of a galaxy: the \citet{Viaene2017} curve for NGC~5626, which was derived with the same method (Fig.~\ref{fig:globalAtt}, green line). It is immediately evident that FCC~167 has a \textit{flatter} curve than NGC~5626. This could point to different dust mixes or, more prosaically, to a different dust geometry. While both dust lanes can be traced back to a ring of dust (see also Sect.~\ref{sec:RTmodel}), in NGC~5626 the dust lane has a lower inclination ($\sim 45 \degree$) than FCC~167 ($\sim 77 \degree$). Steeper curves can be found when relatively more stars are behind the dust lane \citep{Witt2000, Viaene2017}. One could thus naively expect that, at a given dust mix, FCC~167 produces a \textit{steeper} attenuation curve, which is not the case. Alternatively, flatter attenuation curves could point to -- on average -- larger grains, or, a lack of smaller grains \citep{Jones2013}. These considerations thus suggests that the small grains were destroyed in FCC~167. Indeed, it is possible that the survival time of small grains is shorter in the center of a galaxy, than in large-scale dust lanes at larger galactocentric radii such as in NGC~5626.

In Fig.~\ref{fig:globalAtt}, we also highlight the wavelength of several important emission lines (shaded green) and of the NaD absorption line (shaded purple). The emission lines all show a dip in the attenuation curve. This was also noted by \citet{Viaene2017} for NGC~5626 and attributed to spatial asymmetries in our nearly monochromatic images due the emission-line flux. At these wavelengths, the assumption of a symmetric and predictable brightness profile reaches its limit and the intrinsic flux behind the dust lane is underestimated. On the other hand, the NaD lines can also signal additional interstellar absorption due to cold-gas material (in which neutral Sodium can exist, see e.g. \citealt{Davis2012,Sarzi2017,Nedelchev2017}). This may well be the reason for the enhanced attenuation at these wavelengths in the attenuation curve of NGC~5626. Conversely, the absence of such a NaD excess in FCC~167 means that the NaD absorption lines in the MUSE spectra are purely photospheric in origin.

We can also compare these two empirically measured, spectrally resolved attenuation curves with the classic average extragalactic attenuation law from \citet{Calzetti2000}. Their method is based on the modeling global measurements of the UV slope and Balmer decrement in star-forming galaxies, and was recently applied on a large set of SDSS galaxies by \citet{Battisti2016, Battisti2017}. Both attenuation curves are similar (Fig.~\ref{fig:globalAtt}, blue lines). In addition, we plot the extinction law of the THEMIS dust model for Milky Way dust \citep{Ysard2015, Jones2017}. It is important to note that this is in fact an extinction law and not an attenuation curve. Nevertheless, normalised to the $V$ band central wavelength, an extinction law mimics the effect of a thin foreground screen of dust, which is also the implicit assumption in the Calzetti method. The THEMIS curve also lies close to the Calzetti and Battisti curves in the optical (although they diverge in the UV spectrum). These literature curves seem to find the middle ground between the steeper NGC~5626 curve an the flatter FCC~167 curve. From this sparse comparison sample, one could see the validity of Calzetti-like attenuation curves as an average attenuation law. However, users should be aware that large differences can occur on the level of individual galaxies.

\begin{figure}
	\includegraphics[width=0.5\textwidth]{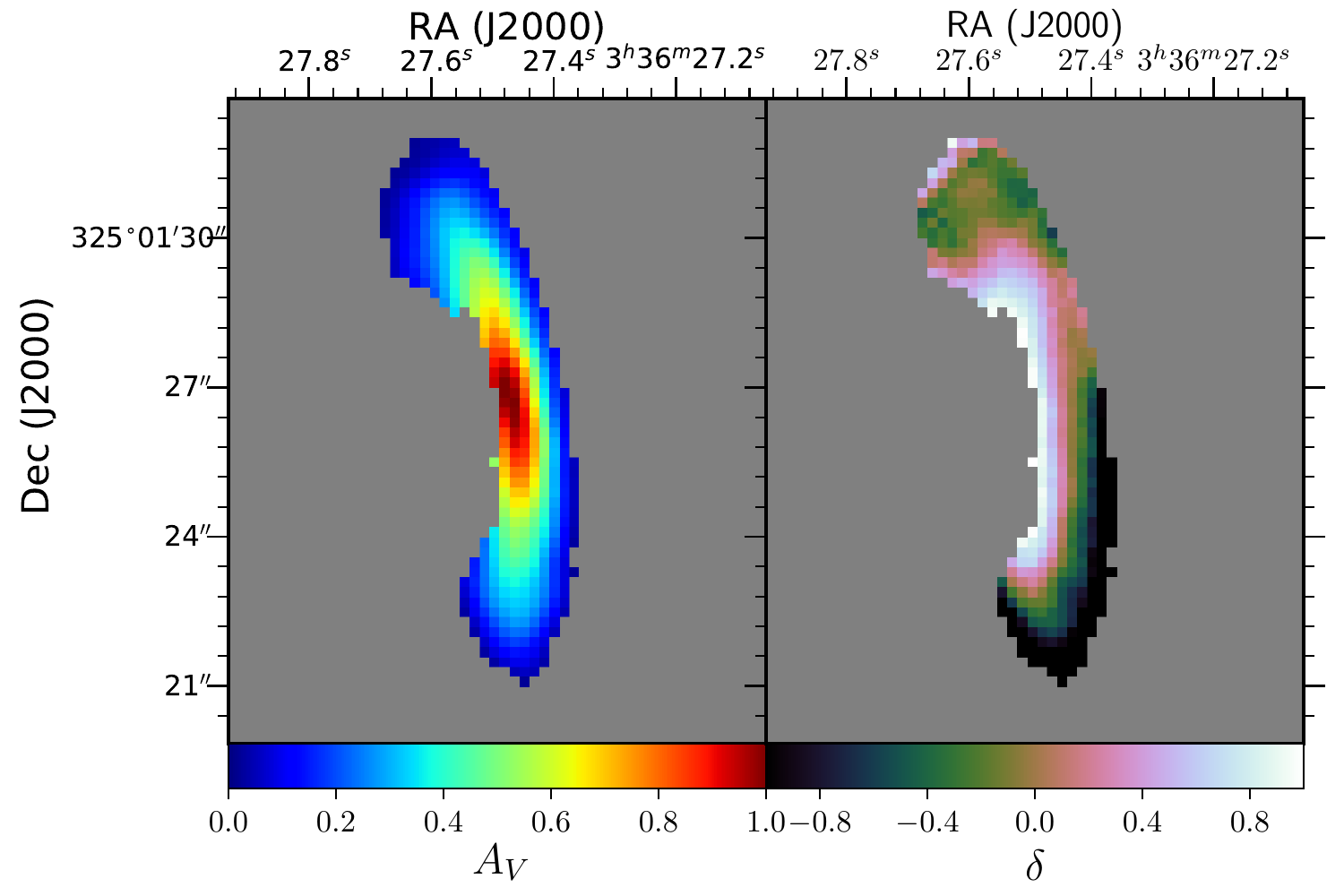}
    \caption{Parameter maps for the fitted pixel-by-pixel attenuation curves. Left: $V$-band attenuation $A_V$. Right: slope parameter $\delta$, where steeper slopes are associated with more negative $\delta$ and a Calzetti-like attenuation curve has $\delta=0$.}
    \label{fig:AttMap}
\end{figure}

The maps of $A_\lambda$ for every wavelength bin allow us to investigate the attenuation curve on a pixel-by-pixel basis in the dust lane of FCC~167. We focus on two main properties here: the slope and the strength of the attenuation. To measure this, we fit a generalised Calzetti law to the attenuation curve of each pixel:
\begin{equation} \label{eq:powerlaw}
A_\lambda = A_V \left( \frac{k(\lambda)}{k_V}\right) \left( \frac{\lambda}{0.55\, \mu\mathrm{m}}\right)^\delta, 
\end{equation}
where $k(\lambda)$ is the \citet{Calzetti2000} curve, and $A_V$ and $\delta$ can change the shape of $k(\lambda)$ by varying the strength and slope, respectively. The parameter maps for $A_V$ and $\delta$ are shown in Fig.~\ref {fig:AttMap}.

The model $A_V$ map shows a regular morphology and closely resembles the attenuation map at $5509\, \AA$ in Fig.~\ref{fig:MGEfit}. This is expected and validates the fitting of Eq.~\ref{eq:powerlaw} to each pixel independently. The morphology of $A_V$ also suggests a ring (rather than e.g. an exponential disk) of dust as the 3D structure behind the dust lane. Interestingly is the slope map ($\delta$, right panel in Fig.~\ref{fig:AttMap}), which has high values (flatter slopes) along the East-side edge and the lowest values (steepest slopes) towards the South and South-West edges. The middle regions of the dust lane have Calzetti-like attenuation curves ($\delta \approx 0$). Naively, one could interpret this as variations in the extinction curve of the dust, and thus as variations in the dust grain size distributions. However, the change in $\delta$ is particularly smooth while dust mix variations are expected to be more localised and associated to clumpy dust structures. In addition, this assumes that the star-dust distribution along the line-of-sight is the same in each pixel, which is unlikely to be the case.

\citet{Viaene2017} investigated the geometric effects of a thick dust slab embedded in a stellar body. They concluded that steep slopes (lower/negative $\delta$) are caused by relatively more stars lying behind the dust slab along the line-of-sight. Vice versa, flatter slopes (higher/positive $\delta$) can be explained by a dust structure which lies further away and is deeper embedded in the stellar body. In this interpretation the East side of the dust lane lies further away than the West side. This is consistent with an inclined dust ring as there is also no visible dust lane on the East side of the galaxy. 

In addition, the difference in $\delta$ would also imply that the North side of the dust lane must lie further away than the South side. For an axisymmetric ring of dust of uniform density, the measured $A_V$ should then be lower on the North side. This is however not what we observe. Unfortunately, within this simple framework we can not make the distinction between a possible dust density gradient from North to South, or changing dust grain properties.

\section{ISM gas properties} \label{sec:ISM}

\subsection{Ionised gas}

\begin{figure*}
	\includegraphics[width=1.0\textwidth]{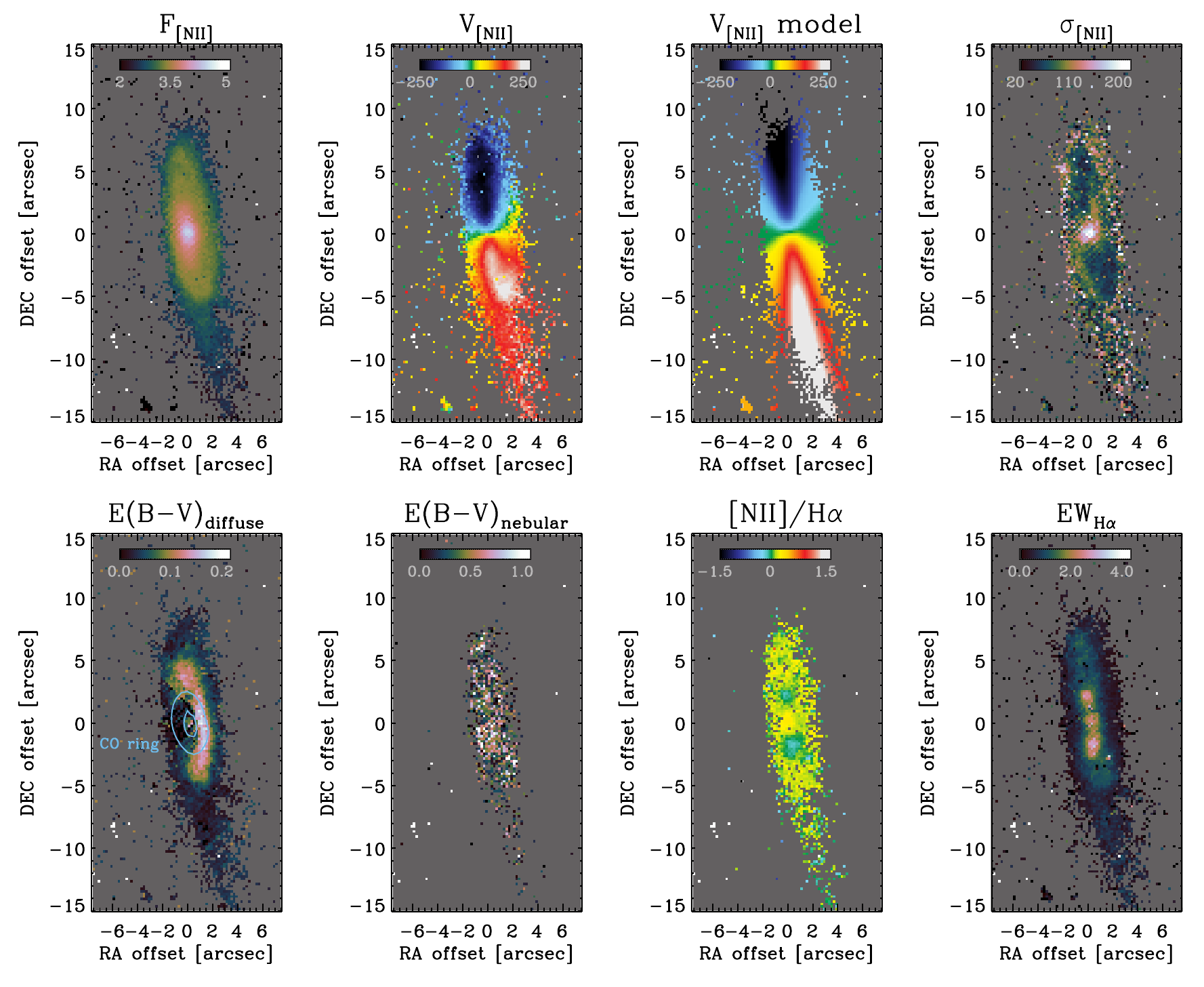}
    \caption{The ionised gas in FCC~167. Top left: surface brightness (in log scale) of the [N{\sc ii}] $\lambda6583$ line highlighting already a tail of ionised gas connecting to the main reservoir. Top second: velocity map indicating organised motion along the major axis, with several additional kinematic features towards the outskirts. Top third: an axisymmetric kinematic model of the measured velocity map. Top right: velocity dispersion of the [N{\sc ii}] line showing a central peak and hints of a spiral-like shape. Bottom left: reddening, $E(B-V)$, by diffuse dust (as opposed to emission-line reddening) derived from the spectral fitting. Blue contours correspond to molecular gas emission measured by ALMA (see Fig.~\ref{fig:ALMA}). Bottom second: reddening, $E(B-V)$, of the emission lines which is much stronger than the diffuse reddening component, and without any distinct features. Bottom third and right panels show the [N{\sc ii}]/H$\alpha$ line ratio  and the H$\alpha$ EW, respectively. Together these two maps indicate low level star formation inside the dust ring and nuclear ionisation due to an AGN.}
    \label{fig:ionisedGas}
\end{figure*}

Dust and gas are usually associated with each other in galaxies. The MUSE spectra allow us to study the ionised gas through emission lines associated with this component. The F3D analysis pipeline produces a
pPXF \citep{Cappellari2004,Cappellari2017} and GandALF \citep{Sarzi2006} best-fit spectral model for each spatial bin. The spatial bins are determined by Voronoi binning to achieve similar signal-to-noise in the stellar continuum. This already provides resolved maps of the ionised gas properties. 

However, to further push in spatial resolution we proceed to fit each individual MUSE spectrum with GandALF while adopting the stellar kinematics and the best-fitting stellar population model returned by pPXF and GandALF, respectively, for the Voronoi bins that contain the MUSE spaxel that is being re-fitted \citep[see also][]{Sarzi2018}.
The results are high-quality measurements of the line fluxes and velocity moments in each spaxel. The only limiting factor is now the instrumental spatial resolution of $70$ pc.

Our main ionised-gas measurements and results are shown in Fig.~\ref{fig:ionisedGas}. Starting with the ionised-gas surface brightness (top left panel), the map for the observed flux of [N{\sc ii}]~$\lambda6583$ line traces the smooth distribution of an inclined disk of ionised gas with a radially declining flux. The disk structure extends beyond the contours of the dust lane, and a remarkable tail extends from the disk in the South-West direction.

The ionised gas reveals clear and regular rotation (Fig.~\ref{fig:ionisedGas}, top second panel), with speeds of around $250$ km s$^{-1}$ in the main body. A small twist is visible in the velocity minor axis towards the outer regions of the disk. This is consistent with a spiral-like shape in the velocity dispersion map (top last panel). The velocity dispersion is further enhanced in the center of the galaxy, due mostly  to beam smearing but possibly also owning to shocks or kinetic energy input associated to a central AGN activity.

\begin{figure*}
    \begin{subfigure}[b]{0.37\textwidth}
        \includegraphics[width=\textwidth]{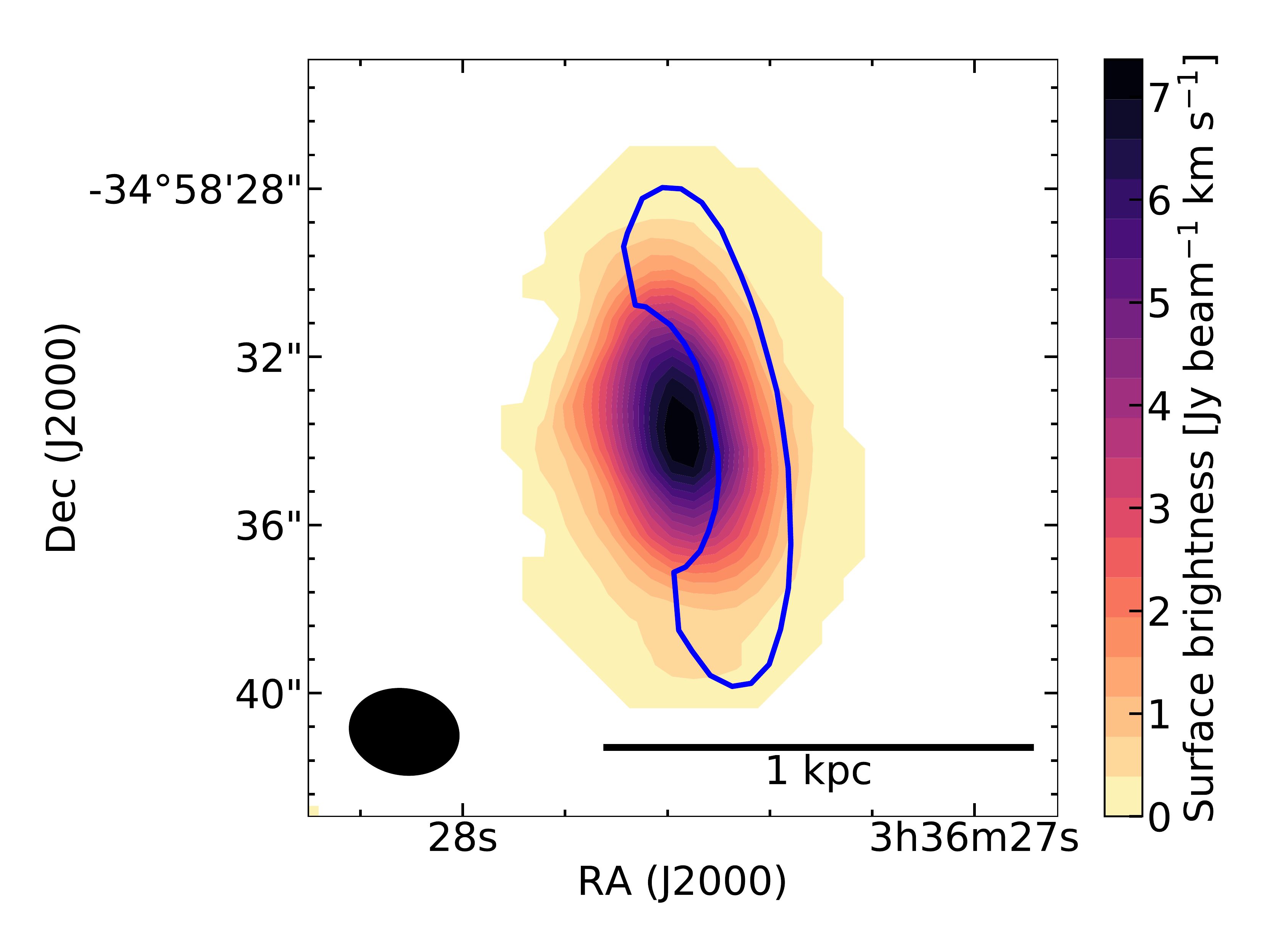}
    \end{subfigure}
    \begin{subfigure}[b]{0.3\textwidth}
        \includegraphics[width=\textwidth]{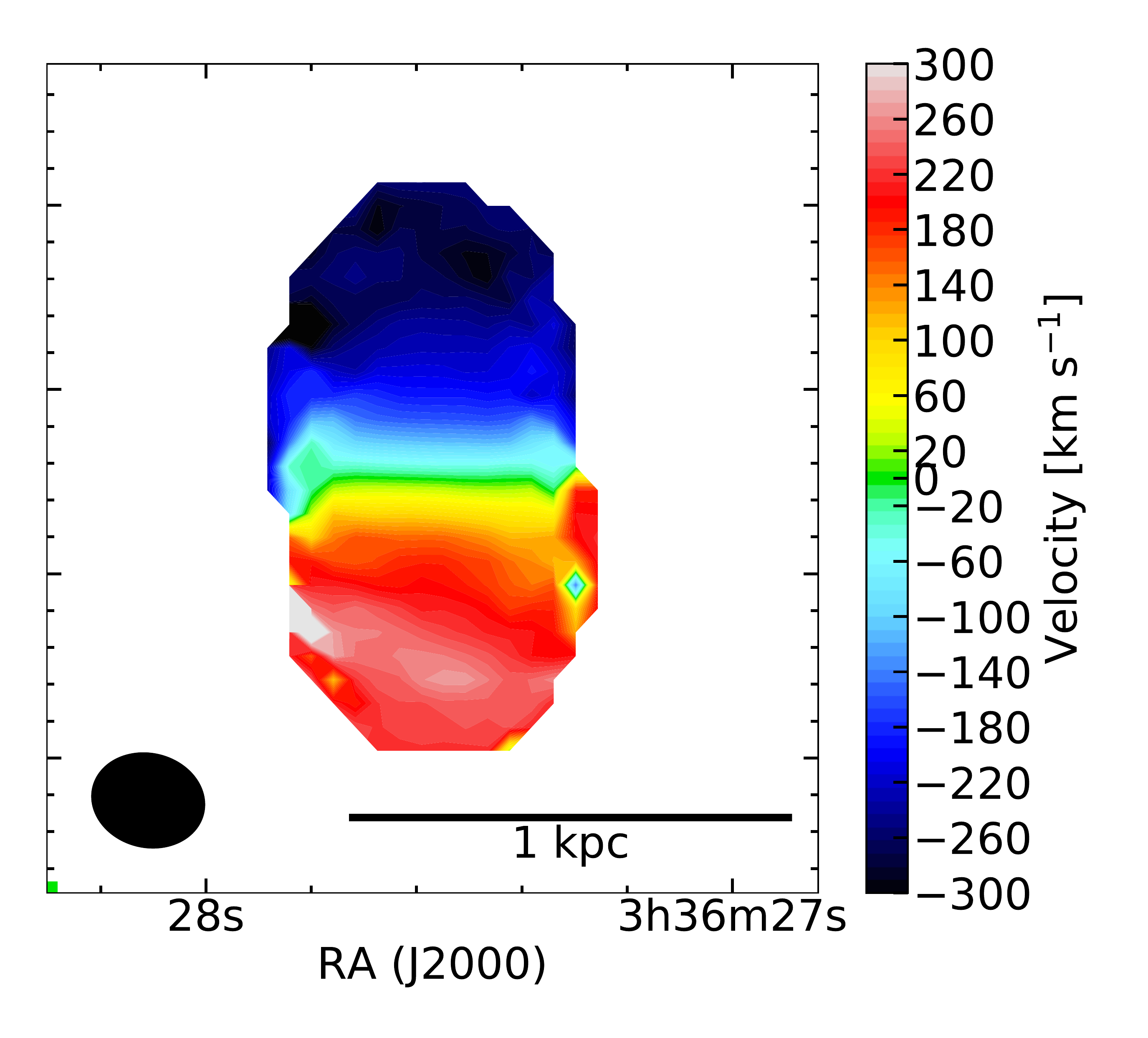}
    \end{subfigure}
    \begin{subfigure}[b]{0.3\textwidth}
        \includegraphics[width=\textwidth]{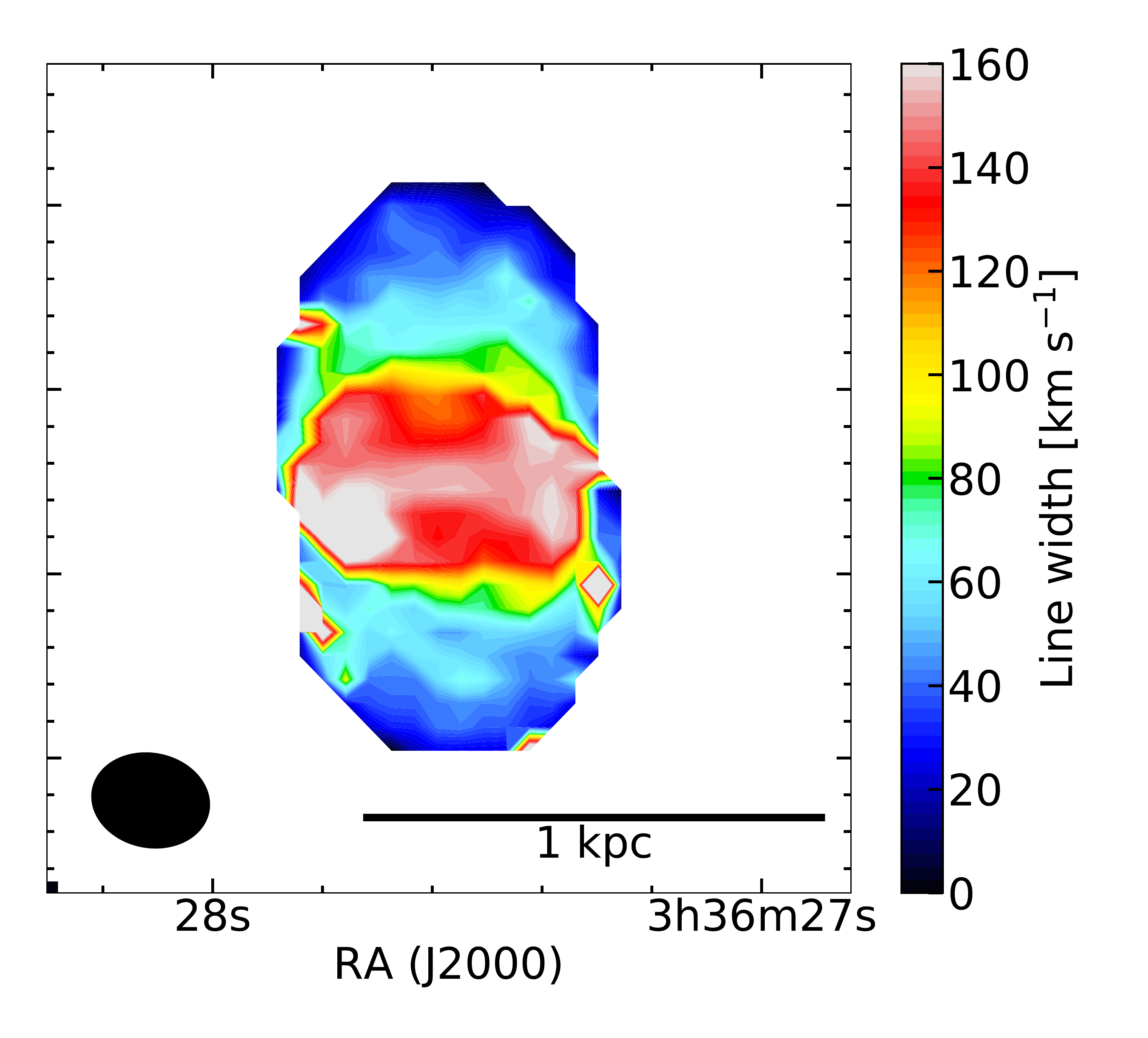}
    \end{subfigure}
    \caption{Moment maps of the CO(1-0) line observations with ALMA. Left: total line intensity with the region of the dust lane overlaid. Middle: velocity map indicating a regular rotation. Right: linewidth map (which peaks in the centre due to beam smearing). The size of the ALMA beam is shown in each frame as a black ellipse.}
    \label{fig:ALMA}
\end{figure*}

Fig.~\ref{fig:ionisedGas} also shows a preliminary axisymmetric model for the ionised-gas kinematics (top third panel). This model was constructed using the KINematic Molecular Simulation (KinMS, \citealt{Davis2013}) package, which can be used to simulate arbitrary gas distributions. The stellar gravitational potential is based on our best MGE model and a mass-follow-light assumption (see also Sect.~\ref{fig:ALMA}). Whereas the model is not perfect, we can already use this to highlight the higher velocity values in the South compared to the observations. From this we conclude that the tail of ionised gas is blueshifted with respect to the main ionised gas disk.

One of the benefits of the GandALF fitting procedure is a simultaneous fit of dust reddening in the lines, and reddening of the stellar continuum emission. The bottom left panel of Fig.~\ref{fig:ionisedGas} shows the reddening by diffuse dust (i.e. excluding Balmer decrement line reddening) as quantified by the $E(B-V)$ metric. This map bears clear morphological resemblance with the $A_V$ map in Fig.~\ref{fig:AttMap}, and the dust lane in Fig.~\ref{fig:MGEfit}. For diffuse dust, stronger reddening coincides with higher attenuation suggesting - to first order - that the line-of-sight geometry and dust grain properties are similar throughout the dust lane. 

In contrast, the distribution of the line reddening (bottom second panel in Fig.~\ref{fig:ionisedGas}) does not follow a ring shape. It is more uniformly distributed across the dust lane region. Reddening in both components in the tail is marginally detected indicating only low levels of dust there. Interestingly, the line reddening is roughly six times higher than the reddening by diffuse dust across the ISM. The sources of the ionised gas are thus much more embedded than the main stellar body. We will further discuss the implications of diffuse vs. clumpy ISM on the observations in Sect.~\ref{sec:discussion}.

The last two bottom panels of  Fig.~\ref{fig:ionisedGas}, showing maps for the [N{\sc ii}]$\lambda6583$/H$\alpha$ ratio and the equivalent width (EW) of H$\alpha$, convey information on the sources of ionization at work in FCC~167. A standard BPT classification based on the [N{\sc ii}]/H$\alpha$ vs. [O{\sc iii}]$\lambda5007$/H$\beta$ diagram \citep{BPT,Veilleux1987} and the division lines drawn and adopted by \citet{Schawinski2007} reveals that the ionised-gas emission falls in the LINER-like category everywhere. The exception are the regions $3^{\prime\prime}$ away from the center that corresponds to the local minima in the [N{\sc ii}]/H$\alpha$ map, just inside the dust-lane regions. Here the nebular emission is classified as composite and is likely powered by some combination of low-level star formation and the radiation from either old stars or a central AGN emission. The EW of Ha map confirms the presence of all three such ionizing sources. This indeed shows both rather constant EW values as expected in the case of LINER-like emission powered by post-AGB stars \citep{Binette1994, Stasinska2008, Sarzi2010} and three distinct peaks that point to the additional ionizing sources. The central peak is due to a true LINER AGN and the other two owing to on-going star formation as their location corresponds to the minima in the [N{\sc ii}]/H$\alpha$ map.

The dominant morphology in the $E(B-V)$ map corresponds remarkably well with the derived $A_V$ maps in figures \ref{fig:MGEfit} and \ref{fig:AttMap}. This again suggests a ring-like structure for the dust, even though this is not obvious from the ionised gas surface density. Indeed, one has to note that the observed nebular emission only traces gas that is ionised and not the total hydrogen reservoir. The fraction of gas that is ionised obviously depends on the density of ionizing photons. In this case of FCC~167 this will correlate with the main stellar body, which is centrally peaked, where ionisation is due to old stars. In addition there is the source of ionizing photons from the low-luminosity AGN in the center and from any OB stars being formed. The result is a stronger ionization field in the center, which explains why the [N{\sc ii}] line (Fig.~\ref{fig:ionisedGas}, top left panel) looks more like a centrally peaked disk of material. Despite these caveats, it is still possible that the ionised gas is tracing a centrally peaked gaseous disk rather than a ring, and that the dust distribution looks more like a ring due to AGN ionisation keeping the disk in the centre too hot for dust or molecules to survive. It is is therefore mindful to also explore the molecular gas in FCC~167.

\subsection{Molecular gas}

The ALMA Fornax Cluster Survey (PI: T. A. Davis) mapped the molecular gas in FCC~167 with ALMA \citep{Zabel2018}. The survey targets the CO(1-0) line and the continuum emission around 3 mm for all galaxies in the Fornax cluster with a stellar mass $> 3 \times 10^8 \, M_\odot$ that contain dust \citep{Fuller2014} or HI (Serra et al. in prep.). Fig.~\ref{fig:ALMA} shows the CO(1-0) line observations for FCC~167. 

There is strong, extended CO emission from the center of FCC~167 which overlaps with the dust lane area (Fig.~\ref{fig:ALMA}, left panel). It also extends towards the East, where the other side of the dust ring resides, but does not cause significant optical attenuation. The ALMA morphology is broader than the dust or ionised gas maps due to the $\sim4^{\prime\prime}$ beam. The velocity and width of the CO line also can be seen in Fig.~\ref{fig:ALMA} (middle and right panels). The molecular gas exhibits regular rotation, with line-of-sight rotational velocities up to $250$ km s$^{-1}$. Typical line widths are between $40-160$ km s$^{-1}$ with the highest values in the center due to beam smearing.

We constructed a dynamical model for the molecular gas again using KinMS \citep{Davis2013}. The output of KinMS is a three-dimensional RA-Dec-velocity data cube, that can be compared directly with the ALMA data cube. We use Markov chain Monte Carlo (MCMC) methods, in particular the \texttt{adamet} Python package of \citet{Cappellari2013}, which implements the Adaptive Metropolis algorithm of \citet{Haario2001}, to constrain the morphology of the molecular gas in the centre of the galaxy. Both exponential disks and ring morphologies were fitted to the data, only a ring-shaped model results in a good fit. The best fit results for the various morphological parameters are listed in Table~\ref{tab:RTmodels}. The inclination was fixed at $76.3 \degree$, based on our models of the dust lane (Sect.~\ref{sec:RTmodel}). 

Boizelle et al. (2017) mapped the CO(2-1) transition in the centre of FCC 167, also using ALMA. Their high resolution observations suggest a gravitationally stable and slightly warped disk of molecular gas, with a brighter ring/tightly wrapped spiral structure superimposed in the inner region. The ring morphology fitted to the CO(1-0) emission is fully consistent with that seen in CO(2-1), although the latter appears to be somewhat more centrally concentrated. This may suggest a central depression in the low density CO(1-0) material, that is perhaps carved by the central AGN \citep[see e.g.][]{Davis2018, Trani2018}. 

With increasing distance from the galaxy's center, it seems that the different ISM components are distributed in a nested structure of increasing spatial extent. Most of the molecular gas resides within the dust lane radius. Beyond that, the molecular gas surface density drops, suggesting that the ISM coincident with the more extended gas disk is primarily atomic, and around this is a skin of ionised material.

\section{3D radiative transfer models} \label{sec:RTmodel}

In this section, we set out to model the 3D distribution of stars and dust in FCC~167 with the aim to reproduce the attenuation curve in the dust lane. Dust continuum radiative transfer simulations allow a proper treatment of scattering and non-local absorption. They have been widely used to model broad-band images of galaxies \citep[see][for a review]{Steinacker2013} and study attenuation and dust heating sources in detail \citep{DeLooze2014,Viaene2017a}. Solving the inverse radiative transfer problem is of particular interest here, as it finds the best parametric model given a set of observations. This technique was pioneered by \citet{DeGeyter2013, DeGeyter2014} with their code FitSKIRT, an optimizer for the radiative transfer code SKIRT \citep{Baes2011,Camps2015}. FitSKIRT simultaneously fits radiative transfer models to a series of images (i.e. oligochromatic fitting), and so breaks the degeneracy between dust reddening and colour gradients in the underlying stellar populations. Its optimization scheme is based on genetic algorithms, which are well suited to compare noisy models (from the Monte Carlo radiative transfer) to noisy data. This method has been successfully applied to edge-on galaxies of various types \citep{DeGeyter2014, Viaene2015, Mosenkov2018}. 

\subsection{FitSKIRT ring model}

We use FitSKIRT to find the best fitting dust ring model as the evidence in the above sections support this geometry. We explored the possibility of an exponential dust disk, but this provided no good fits to the data. To constrain the model, we select five image slices from the MUSE datacube, at $4602$, $5509$, $6509$, $8059$ and $9340 \, \AA$. The choice of the slices is meant to span the widest range in wavelength, avoid strong emission or absorption lines, and provide a roughly equal spacing in the wavelength domain. The number of slices is a trade-off between sampling the wavelength range and computational efficiency. We found that with these five images the models converge well towards an optimal solution. Each fit is run ten times with a different random seed to assess the uncertainty in the best-fit parameters.

To obtain a parametrisation for the stellar body, we first fit a MGE model to the integrated light image of FCC~167, masking out the dust lane. This combination of 2D Gaussians can be easily deprojected to a 3D profile assuming an inclination angle, which is a free parameter. The MGE model describes the overall shape of the stellar body, and the relative brightness ratios of the different Gaussian functions. The total luminosities (at each wavelength) of the stellar component are left as free parameters. This approach makes no assumptions about the underlying stellar populations and their associated colours, and limits the number of free parameters in the stellar component.

The ring of dust is described by an axisymmetric density profile in cylindrical coordinates $r$ and $z$. It is defined by a Gaussian with radius $R$ and FWHM $W$, and a vertical scale height $H$, all of which are free parameters:
\begin{equation} \label{eq:dustRing}
\rho_d(r,z) = \rho_0\,\exp\left[ -\frac{4\ln{2}(r-R)^2}{W^2} \right] \exp\left( -\frac{|z|}{H} \right).
\end{equation}
The FWHM $W$ is trivially related to the Gaussian standard deviation $\sigma$ = FWHM$/(2\sqrt{2\ln{2}})$. $\rho_0$ is the central density and also left as a free parameter linked to the total dust mass. Thus, there are five free geometric parameters, and five free luminosity parameters. This is a modest task for FitSKIRT, which was tested on fitting more complex edge-on spiral galaxies \citep{DeGeyter2014}, with up to 11 geometric parameters and only 4 luminosity constraints (the SDSS $griz$ bands, which are comparable to the wavelength range available here). We should therefore be able to find the best configuration, within the limits of the chosen geometry (a ring). 

\begin{figure*}[h!]
	\includegraphics[width=\textwidth]{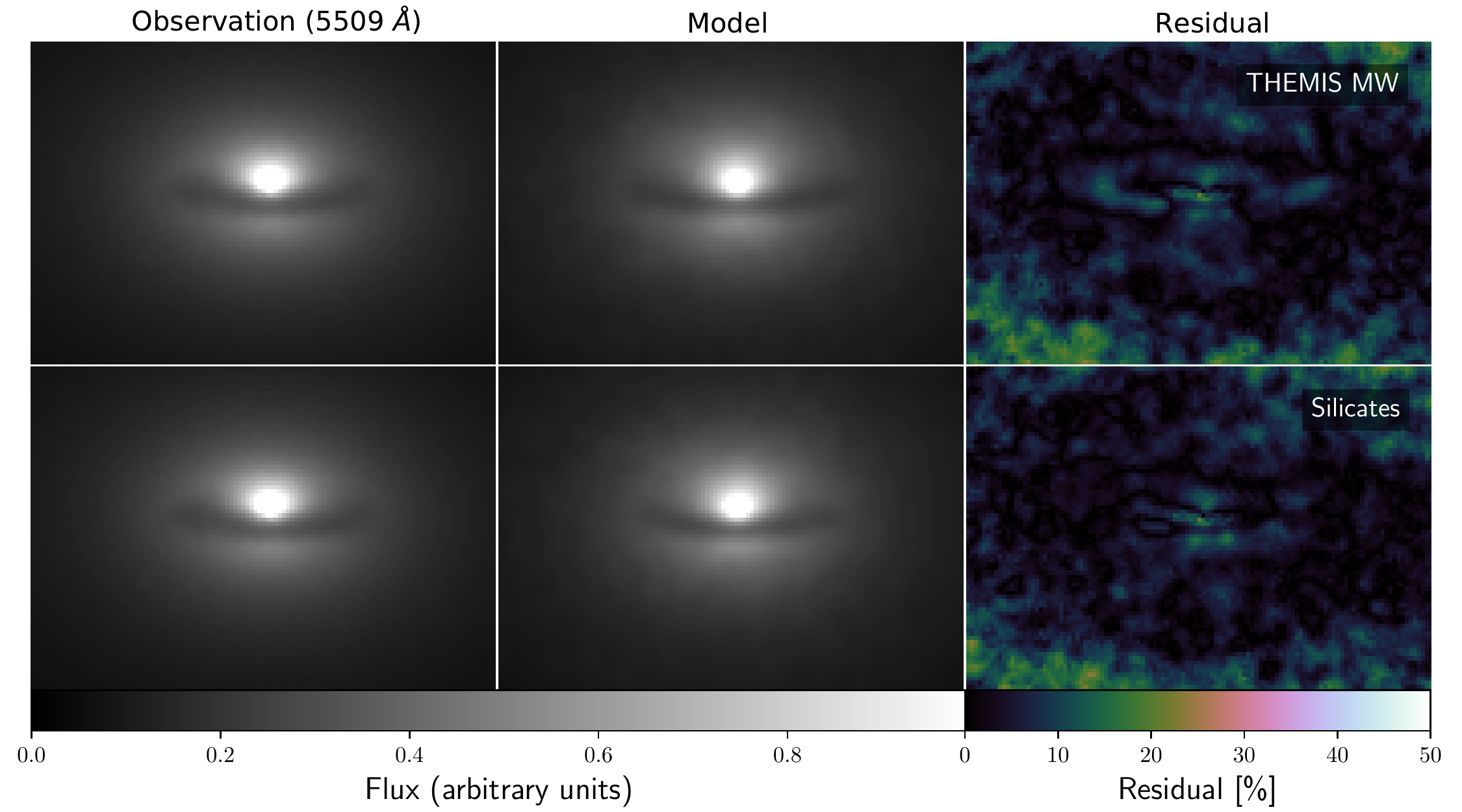}
    \caption{Best fit models for a ring of dust with a THEMIS Milky Way dust mix (top row) and for a THEMIS bare sillicate dust mix (bottom row). We only show a representative image slice at the $V$-band central wavelength ($5509\, \AA$) as the results are very similar across the spectrum. Both observed images in the first column are the same. The second column shows the corresponding model images and the third column shows the residual maps following $\frac{I_\text{obs}-I_\text{mod}}{I_\text{obs}}$ in pixel flux.}
    \label{fig:FitSKIRTring}
\end{figure*}

In Fig.~\ref{fig:FitSKIRTring} we present the model for an example image slice, at the $V$-band central wavelength ($5509 \, \AA$). The results and model quality are very similar across the spectrum. In this first model, we use the THEMIS mix for diffuse Milky Way dust \citep{Ysard2015, Jones2017}. Visually, the model images resemble the observed ones remarkably well. Monte Carlo noise and PSF convolution of the model images leads to a basic level of noise in the model maps. Adding to that the inevitable observational noise, it becomes speculative to interpret structures below the $10\%$ residual level. Still, small deviations can be spotted in the residual maps, which are mostly associated with the densest parts of the dust ring. The model underestimates the attenuation in these regions by $10-20\%$. In addition there is a discrepancy in the central zone, right above the dust lane: the model pixel values are $15-25\%$ lower than the observed ones. This could be an overestimate of the attenuation in the inner parts of the dust ring, and could be a misfit in the stellar body. In the latter case, the MGE model is not completely capable to match the brightest pixels in the core of FCC~167. This may be related to the presence of a low-luminosity AGN. Nevertheless, we consider this to be a very good radiative transfer fit as model deviations lie well below $25 \%$.

\begin{table*} 
\caption{Best-fit model parameters for the ISM in FCC~167. Top row corresponds to the molecular ring, where the inclination was kept fixed. Bottom rows correspond to the dust ring geometry. Two different dust mixes were run: a canonical THEMIS Milky Way dust mix and a dust mix with only bare silicate grains. Dust mass and inclination are free parameters, as well as the dust ring's dimensions. In all cases, the geometry is defined by Eq.~\ref{eq:dustRing}.}
\label{tab:RTmodels}
\centering     
\begin{tabular}{lccccc}
\hline
 Model & $\log(M/M_\odot)$ & radius $R$ (pc) & width $W$ (pc)  & height $H$ (pc) & Inclination $i$ (degrees) \\
\hline
ALMA CO & $7.67 \pm 0.06$ & $151 \pm 62$ & $223 \pm 12$ & $223 \pm 8$ & $76.3$ (fixed) \\
\hline
THEMIS MW & $4.70 \pm 0.03$ & $321 \pm 40$ & $276 \pm 99$ & $16.6 \pm 6.0$ & $77.6 \pm 1.8$ \\
Silicates         & $4.80 \pm 0.04$ & $285 \pm 53$ & $217 \pm 151$  & $13.6 \pm 8.2$ & $76.3 \pm 1.2$ \\
\hline
\end{tabular}
\end{table*}

Our second model starts from the exact same geometry, but relies on a stripped version of the dust mix. To assert maximum contrast, we removed all carbonaceous material from the THEMIS Milky Way model. There are no amorphous hydrocarbons as small grains (the most significant alteration), and the large silicate grains are stripped from their carbon-rich mantles, leaving only the bare silicate grains. The average grain size of this dust mix is larger, which will further affect the attenuation and emission properties. The best-fit model for this dust mix is shown in the bottom row of Fig.~\ref{fig:FitSKIRTring}. Visually, this model looks almost identical to the first one and matches the observations remarkably well. The residual maps are slightly better with most pixels below $20\%$ and only a few in the $20-25\%$ range. The largest deviations can again be associated to the nuclear regions, where the MGE component seems to fall short to explain the sharp rise in brightness towards the central pixels. \\

The best-fit parameters and their uncertainties are summarised in Table~\ref{tab:RTmodels}. The uncertainties were derived from running the fit 10 times and computing the standard deviation on the spread in each parameter. The two models produce marginally consistent dust masses of $(5.01 \pm 0.34) \times 10^{4} M_\odot$ vs. $(6.30 \pm 0.60) \times 10^{4} M_\odot$, with a higher mass for the silicate model . Both models also retrieve the same inclination angle ($77.6\degree \pm 1.8\degree$ and $76.3\degree \pm 1.2\degree$ for the MW and silicate models, respectively). The inclination angle is perfectly consistent with the value of $77.6\degree$ obtained from the stellar dynamical modeling presented in \citet{Sarzi2018}.

The structural parameters for the dust ring are also consistent between the two dust mix models, but the uncertainties are larger. On the one hand, the MW dust model finds a radius of $321 \pm 40$  pc, a FWHM of $276 \pm 99$ pc and a scale height of $16.6 \pm 6.0$ pc. On the other hand, a smaller ($R=285 \pm 53$ pc) and more narrow ($W=217 \pm 151$ pc) ring is found for the silicate model, with a slightly lower scale height ($H=13.6 \pm 8.2$ pc). While the relative uncertainties on the radius are reasonable, they are much higher for the ring width and scale height, especially for the silicate model (up to $70\%$). The uncertain values for the scale height may be linked to the resolution limit of the data. The best-fitting scale heights are comparable with the pixel scale of $17.8$ pc and well below the PSF FWHM of $70$ pc. The dust scale heights are an order of magnitude smaller than the best-fit model for the molecular gas. This can mostly be attributed to the difference in beam size. The molecular scale height corresponds to $0.6$ times the $4^{\prime\prime}$ ALMA beam.

There may also be some weak degeneracies in the fitting;  a dust lane may be represented by a wide ring and smaller radius, or by a larger radius and more narrow ring. A key diagnostic to break this degeneracy is an accurate representation of the dust attenuation in the central regions, and of the peak attenuation in the ring. Both aspects are influenced by the underlying stellar distribution. As we noted before, the MGE model is unable in perfectly retrieving the central peak and the global brightness level at the same time. As such, FitSKIRT can only retrieve the structural parameters of the dust ring up to a certain level of accuracy. Still, the level of accuracy seems reasonable given the complexity of the problem.

\begin{figure}
	\includegraphics[width=0.5\textwidth]{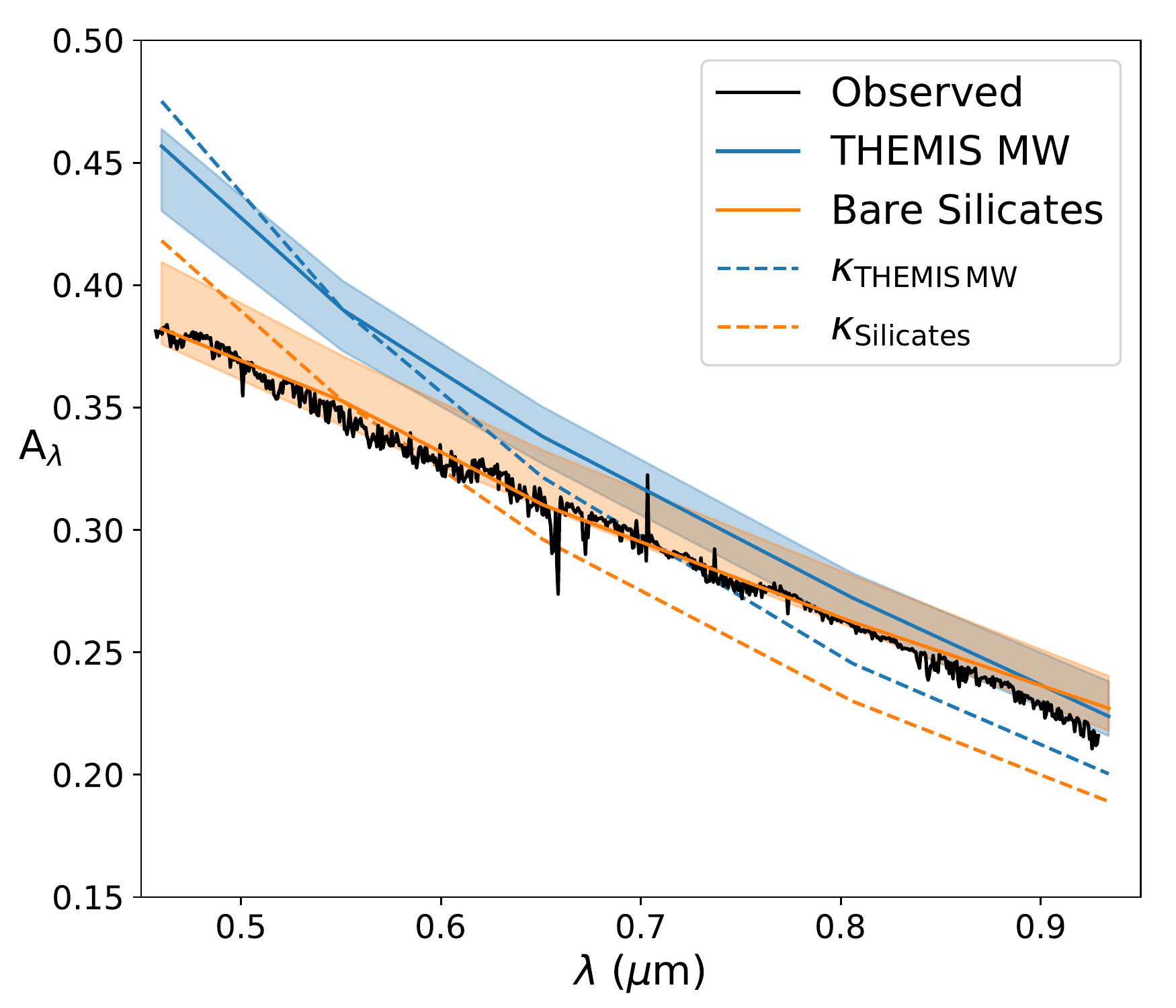}
    \caption{Attenuation curves of the best fit radiative transfer models for a dust ring geometry. The THEMIS Milky Way dust mix produces a steeper curve (blue line) than a dust mix with only bare silicate grains (orange line). The dashed lines correspond to the corresponding intrinsic extinction curves of the dust mixes, normalised to the model attenuation at $5509$ \AA. The observed attenuation curve of FCC~167 (from Sect.~\ref{sec:attenuation}) is shown in black.}
    \label{fig:attenuationCurves}
\end{figure}

An important comparison to differentiate the two FitSKIRT models is their optical attenuation curve. One of the output products of the fitting is a dust-free image at every wavelength bin. We measured the flux in the dust lane area for the model image and the dust-free counterpart and applied Eq.~\ref{eq:att} to obtain the attenuation curve. This can directly be compared to the empirically derived global attenuation curve from Sect.~\ref{sec:attenuation}. The attenuation curves for both dust mixes are shown in Fig.~\ref{fig:attenuationCurves}, along with their model uncertainties. The model with the Milky Way mix overestimates the observed attenuation level by $0.05-0.08$ mag in the blue part of the spectrum, but only by $0 - 0.02$ mag in the red part. The general slope of this model curve is significantly steeper than the observed one. The comparison is much better for the bare silicate model, with offsets consistently between $0-0.03$ mag across the wavelength range. The best-fit attenuation curve for this model perfectly matches the observed slope. This is perhaps the clearest difference between the two radiative transfer models for FCC~167. While Milky Way dust results in an attenuation curve that is clearly too steep, the bare silicate dust mix naturally produces the observed  slope. 

For completeness, we also plot the corresponding intrinsic extinction curves $\kappa$ for each dust mix in Fig.~\ref{fig:attenuationCurves}. We normalise these geometry-free curves to the model attenuation at $5509 \AA$ for ease of comparison. This shows clearly the effect of the ring geometry for both models, converting a steeper extinction curve onto a flatter attenuation curve. Interestingly, the slope difference in extinction is significant, but not that large. This is then amplified by the geometry into a large slope difference in attenuation.

\subsection{Panchromatic model and energy balance} \label{sec:energybalance}

\begin{figure}
	\includegraphics[width=0.5\textwidth]{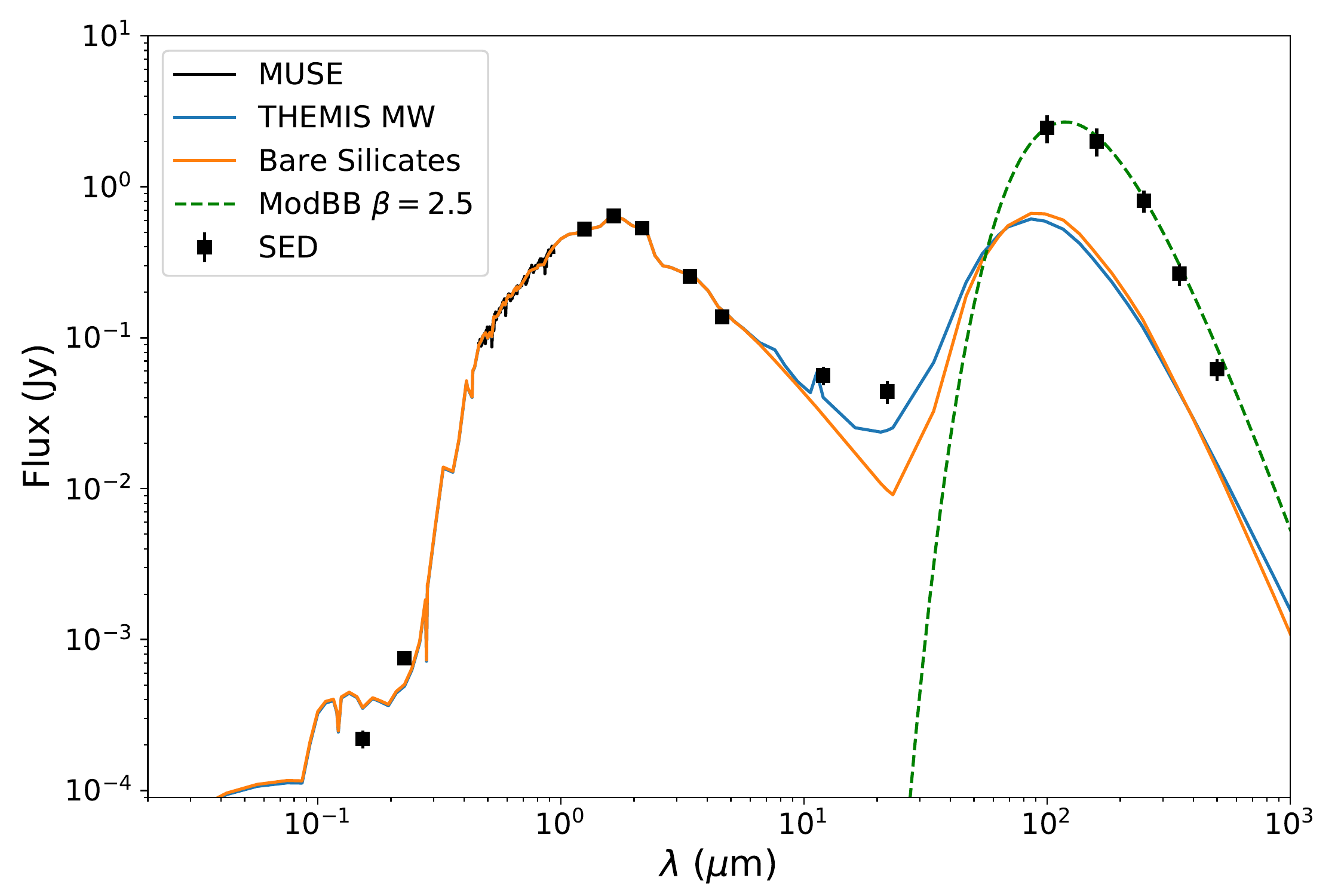}
    \caption{Panchromatic radiative transfer model predictions based on the best-fit FitSKIRT models for THEMIS MW dust and the bare silicate dust mix. A single (10 Gyr) stellar population was used for the intrinsic SED of the stellar component. No additional dust heating sources are adopted in the model. The green line represents a modified black body fit to the \textit{Herschel} fluxes from 100-500 $\mu$m, with a free $\beta$ to match the steep Rayleigh-Jeans slope.}
    \label{fig:PanSEDmodels}
\end{figure}

The center of a massive galaxy like FCC~167 could be a hostile environment for dust grains. The high stellar density results in a strong interstellar radiation field and the $H\alpha$ EW map shows further ionizing radiation from low-level star formation and a LINER AGN. Both stellar sources will contribute to the heating of the dust grains. In addition, there may be other heating mechanisms at work, such as through cosmic rays or collisions with hot gas atoms. To assess the importance of these contributions, we measure the energy deficit in FCC~167. In a scenario where the dust is only heated by the evolved stellar body, there should be a perfect match between the observed FIR/submm flux and the predicted flux from the FitSKIRT model. To simulate the panchromatic emission for each model, we adopt the fitted parameters for the dust component and assign an emission template to the stellar component. To estimate the age and metallicity of the stellar component we used pPXF to fit the MUSE global spectrum extracted in the same central region used for our radiative-transfer modeling. Using the \citet{Bruzual2003} SED templates, this fit converged to $10$-Gyr-old, Solar metallicity average stellar population in the central region of FCC~167. This template was normalised to the total 2MASS $K$-band flux, assuming this band only contains flux from the evolved stellar populations, with no significant dust attenuation. The transfer of radiation is then simulated in 3D with SKIRT. 

Fig.~\ref{fig:PanSEDmodels} compares the panchromatic SED models with the observed fluxes. The fluxes were measured from the DustPedia maps (see Sect.~\ref{sec:data}), and the optical spectrum was extracted from the MUSE datacube. Both models perform well -by design - in the optical and NIR. They also perform reasonably well in the UV, with a slight overprediction of the far-UV radiation. The discrepancies are much larger in the infrared, where neither of the model SEDs reach the observed fluxes. In the mid-infrared (MIR), the bare silicate model lacks any features of transiently heated grains because there are no small grains in the dust mix. The MIR is therefore significantly lower than the MW dust model. In the FIR, both models peak at the same wavelengths, and the silicate model exhibits a slightly steeper Rayleigh-Jeans tail. In fact, the Rayleigh-Jeans slope of the silicate model lies closer to the observed slope than the MW model. But in both cases the dust emission is up to an order-of-magnitude lower than observed, and the dust peaks at shorter wavelengths, indicating a higher temperature.

We calculate the energy deficit by integrating the emission from 12-500 $\mu$m equivalent to the WISE 3 - SPIRE 500 $\mu$m bands. The MW model only produces $35 \%$ of the observed flux, and the silicate model produces $33 \%$. Another way to quantify the energy deficit is to infer the total dust mass from the observed SED. We fitted a modified black body function to the \textit{Herschel} fluxes, with a variable $\beta$ to match the steep Rayleigh-Jeans slope. We find that the opacity to convert flux to dust mass is $\kappa_{350} = 0.348 \, \text{m}^2\text{kg}^{-1}$ for both dust models. Both dust mixes have a similar value because at $350 \, \mu$m (the commonly used wavelength to convert flux to dust mass) the emission is dominated by the large silicate grains. Within the uncertainties, both dust models thus give the same dust mass from the observed fluxes, which is $8.3^{+2.2}_{-1.8} \times 10^5 M_\odot$. This is already more than an order of magnitude higher than the dust masses derived from visual attenuation (Table~\ref{tab:RTmodels}). We discuss this further in the next section.

\section{Discussion} \label{sec:discussion}

\subsection{Dust distribution and survival}

In Sect.~\ref{sec:attenuation} and \ref{sec:RTmodel} we have measured the global optical attenuation curve in FCC~167 and found a radiative transfer model that replicates this attenuation curve. The best match is obtained with a dust mix that contains no small grains. However, the models predict dust emission spectra which lie an order of magnitude below the observed fluxes at long wavelengths. The energy balance problem is widely known in radiative transfer models \citep[see][for the most recent discussion on this topic]{Mosenkov2018} and varies between a factor of 1.5-5 for spiral galaxies. Little research has been done to assess the energy balance in ETGs, but the discrepancy in FCC~167 (a factor of $\sim 3$) is comparable with these studies for spiral galaxies. Several factors seem to be at play here.

First, there could be a diffuse dust component, not associated with the dust lane, but with the general stellar body. This was originally postulated by \citet{Goudfrooij1995} to explain similar order-of-magnitude differences in attenuation-based and emission-based dust masses for ETGs. This theory is supported by \citet{Kaviraj2012} as they find similar mass deficits for a large sample of dust-lane ETGs. However, the most reasonable origin for such a diffuse dust component is mass loss from AGB stars, but mass-loss yields for these stars fall drastically short to explain the observed dust masses \citep{Finkelman2012, Rowlands2014}. \textit{Herschel} observations of FCC~167, while coarse in resolution, show a clear, unresolved FIR/submm peak at the center of the galaxy, and no extended component. Both the ionised gas and the molecular gas are confined to the centre of the galaxy as well. In the optical spectra of FCC~167 we find only weak reddening for the stellar continuum, and the vast majority of it happens inside the dust lane. 

However, evolved stars could also host warm dust in their atmospheres as opposed to dust expelled by winds. Such a dust component would be brightest in the MIR \citep[see e.g.][]{Temi2009}. Using Eq.~1 from \citep{Davis2014} we compute a $3-27 \%$ contribution of this dust to the WISE $22 \,\mu$m flux. The radiative transfer models do not include such warm dust around evolved stars. This could in part explain the $45 \%$ deficit between the observed WISE $22 \,\mu$m flux and the MW dust model flux, and the $78 \%$ deficit for the bare silicate model. Still, the discrepancy is too large to be explained solely by this component.

Second, the dust may be clumpy, which is an explanation that has also been put forward also to rectify the energy balance problem in spiral galaxies \citep{Bianchi2008, Baes2010, DeLooze2012b}. Cold and dense clumps would contribute little to the attenuation of starlight since they are compact. These clumps do have an equilibrium temperature and thus emit FIR/submm light as a modified black body. The difference in dust mass suggests that roughly $7.7\times 10^5 M_\odot$ or $93 \%$ of the observed dust mass should be locked up in clumps. This fraction seems to be particularly high, but it could be a result of intense dust destruction (see below), where only dense clumps can survive. A clumpy medium is also favoured by a much higher line reddening than continuum reddening (Fig.~\ref{fig:ionisedGas}, bottom left panels) and by evidence of opaque molecular clouds in this galaxy \citep[see discussion in Sect. 4.1 of][]{Boizelle2017}. Higher line-to-diffuse reddening ratios are seen in other galaxies, especially for those with higher mass \citep[e.g.][]{Koyama2018}. If line emission originates from compact ionised-gas clouds, then these seem to be much dustier than the diffuse ISM of FCC~167. An additional effect of clumpy dust is a shift of the FIR peak to longer wavelengths as the dust is generally colder. This would bring the current panchromatic models closer to the observed SED, however, we lack the spatial resolution to properly test this in our simulations.

Third, it is likely that the dust is not only heated by the evolved stellar populations. The BPT line ratios reveals low level star formation just inside the dust lane, although the UV and absorption-line signature of young stellar populations may have been lost in the global spectra and broad-band measurements that we extracted in the central regions of FCC~167. Additionally, the large nebular-line reddening suggest that such star-forming regions may be heavily obscured. Ongoing star formation can heat the dust in two ways: a) through conventional photon heating as young stellar populations emit most of their light as energetic UV and optical radiation and b) through generation of cosmic rays which can get caught by dust grains, dissipating energy as heat. 

\begin{figure}
	\includegraphics[width=0.5\textwidth]{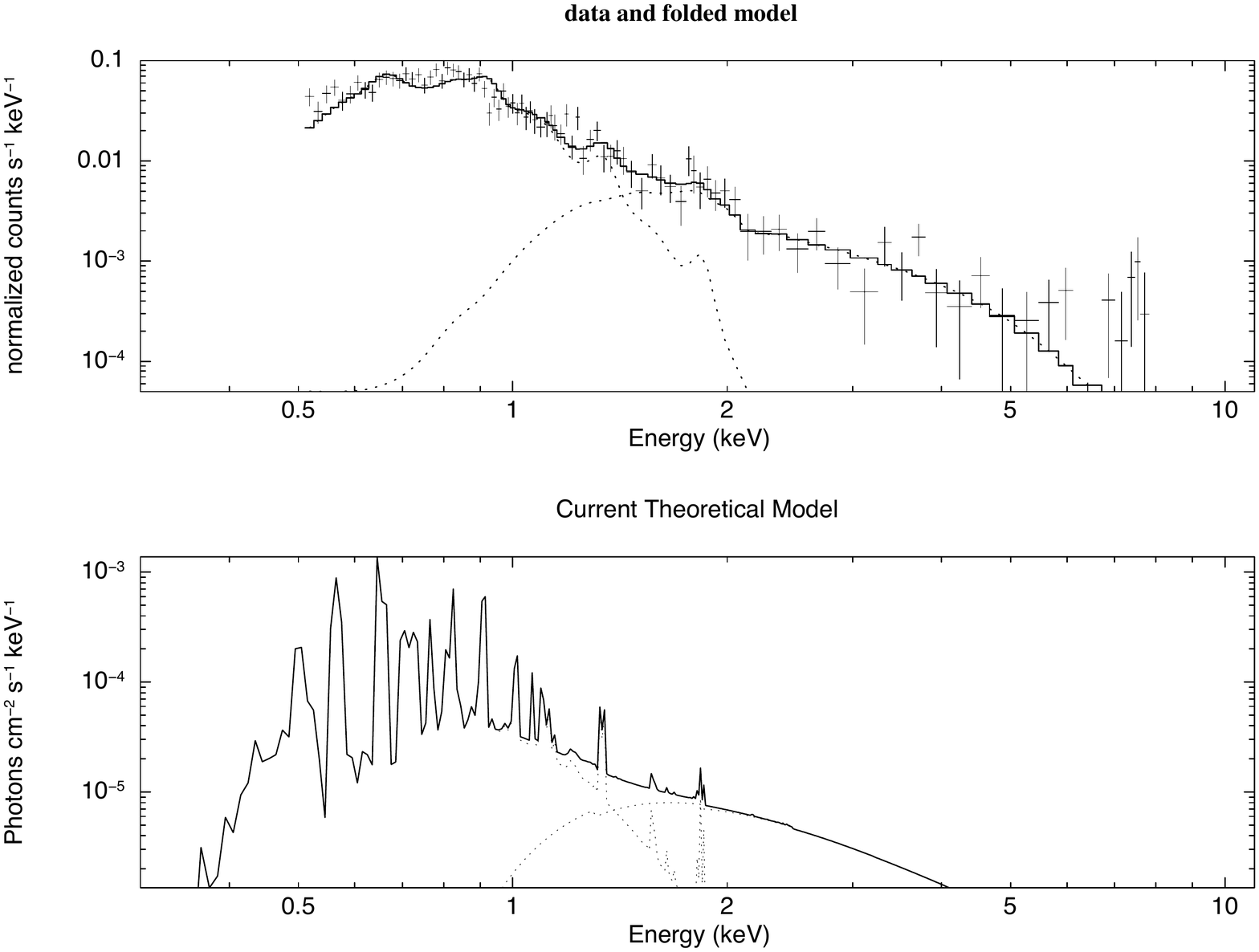}
    \caption{Fit to the Chandra X-ray spectrum of FCC~167. Top: Crosses are the observed photon counts per channel. The black line shows the total model flux. Bottom: unfolded model (solid line) and its separate components (dotted line). A \texttt{wa*po+mekal} model was used to represent a power law and a hot gas line spectrum. The unfolded model is adjusted for spectral resolution and filter response function before comparison to the data in the top panel. }
    \label{fig:xraySpec}
\end{figure}

Finally, it is also not uncommon for high-mass galaxies to contain a hot gas reservoir. When a hot gas electron hits a dust grain, it dissipates away its energy onto the grain, thus heating the dust. An important additional effect is that these electrons can damage the surface of the dust grains. Such `sputtering' by hot gas is an effective dust destruction mechanism and is particularly effective in destroying the small grains \citep{Jones2004}. FCC~167 was detected in one of the fields observed by the \textit{Chandra} X-ray Observatory. The X-ray emission is clearly separated from the main intracluster X-ray emission, which is centered at NGC 1399 \citep{Scharf2005}. In order to assess and separate the relative importance of X-ray binaries and diffuse hot gas to such X-ray emission, we extracted the X-ray spectrum in the nuclear region of the galaxy as shown in Fig.~\ref{fig:xraySpec}.

XSPEC \citep{XSPEC} was used to fit a two-component (\texttt{wa*po+mekal}) model to the observed spectrum. We used an absorbed power-law (\texttt{wa*po}) to account for X-ray photons from unresolved binaries and from the (potential) AGN. To describe the hot gas we use a single-temperature \texttt{mekal} model, which includes line emissions from several elements. The model in Fig.~\ref{fig:xraySpec} shows that a hot gas component is a necessary addition to the power-law continuum. Our fit yields a hot gas density of $7\times10^{-3}$ cm$^{-3}$ , with a temperature of $3.2\times10^7 K$. Dust heating by hot gas could thus be an important factor in FCC~167. 

These hot gas properties can also be used to estimate the dust lifetime following \citet{Jones2004}. Dust grains, fully exposed to the hot gas in this galaxy would be destroyed on a timescale of $0.4$ Myr (very small grains) to $14$ Myr (large grains). At a characteristic galactocentric radius of $285$ pc and rotational velocity of $250$ km s$^{-1}$, the orbital timescale is $\sim7$ Myr, comparable to the dust destruction timescale. The efficient dust destruction in the center of FCC~167 raises the important question of why there is a dust lane in the first place. If the dust lane is stable and long-lived, then it must be self-shielding or partly shielded by the gas. This supports the suggestion that a good fraction of the dust is organised in dense clumps. Ongoing star-formation inside the dust lane will also produce some dust and replenish the current reservoir.

\subsection{Origin of the ISM}

The results acquired in this study allow us to speculate about the origin of the gas and dust in FCC~167. Statistically, dust lanes in ETGs are attributed to gas-rich minor mergers \citep[e.g.][]{Kaviraj2012, vandeVoort2018}. A relic of such a scenario would be the blue-shifted tail of ionised gas falling onto the main reservoir from behind the galaxy. This means we are observing a peculiar moment in the merger history of FCC~167, right after a minor merger. If gas and dust were externally accreted, it will be heavily processed by the hot gas on its way inward. This could explain the lack of small grains we derive from the radiative transfer fit. Once the material settles in a dust ring, it can cool down and start forming stars in the shielded inner regions. In the very centre however, fuelling of the supermassive black hole induces feedback and again dissociates the molecular gas. However, such a minor merger is disfavoured by the fact the galaxy lives in a cluster. Mergers and accretion events are relatively rare in such an environment compared to the field.

Alternatively, we must consider that the gas and dust was already present in the galaxy, and is now being stripped. The gas-to-dust ratio of $\sim80$ in FCC~167 is more in line with disks of spiral galaxies \citep[see Figure 2 in][for a comparison]{Davis2015} and ram-pressure stripping is a common process in a clusters like Fornax. It is however unlikely that FCC~167 is \textit{radially} falling into the cluster center as the tail of ionised gas is pointing towards the central galaxy NGC 1399. To better assess this, we must look at the problem in 3D. Distances based on surface brightness fluctuations measured by \citet{Ferrarese2000} indicate that FCC~167 ($18.8$ Mpc) is closer to us than NGC 1399 ($20.5$ Mpc). However, FCC~167 is moving away from us much faster ($1877$ kms$^{-1}$) than NGC 1399 ($1425$ km$^{-1}$). Thus, based on the radial velocity component, FCC~167 seems to move towards NGC 1399. We can infer the tangential velocity component of the galaxy from the tail of ionised gas, which points South towards NGC 1399 and is blueshifted. The tangential component must point North, so gas is pushed out on the South side and towards us. This would mean that FCC~167 is in fact moving through the cluster around NGC 1399 and is on its way to apocenter. As a consequence, the gas and dust in FCC~167 may be a relic of a much larger reservoir, pre-dating its interaction with the cluster. In the above stripping scenario, the dust gradually gets exposed to the harsh radiation field and can transform into a clumpy, self-shielding medium with little small grains.

\section{Summary and conclusions} \label{sec:conclusions}

The goal of this paper was to address the origin and fate of the nuclear dust lane in FCC~167. To this end we have studied several aspects of the ISM in this ETG. Our investigation was centred around new MUSE observations from the Fornax 3D project. We first measured the attenuation curve in the dust lane area of the galaxy by treating each slice in the MUSE cube as a separate image. A dust-free surface brightness model was compared to the observations to determine the dust attenuation.\\ \textit{We find that the attenuation curve in FCC~167 is quite flat compared to literature curves. Given the high inclination of the galaxy, this is already hard to explain with a canonical Milky Way dust grain mix and points towards a larger average grain size.}

Pixel-to-pixel variations in the attenuation curve strength and slope seem consistent with an inclined ring of dust. To better assess the ISM distribution we also map the ionised and molecular gas in FCC~167. The molecular gas was detected by ALMA in the CO(1-0) line and is more compact than the dust lane. It shows regular rotation and a centrally peaked velocity dispersion. A dynamical model suggests again a ring-like morphology, potentially because nuclear feedback is preventing molecules to form in the very centre of FCC~167. The ionised gas is organised in a regularly rotating disk that extends slightly beyond the dust lane. \\
\textit{The line reddening is roughly $6$ times stronger than reddening of the stellar continuum. Reddening is limited to the area of the dust lane and again consistent with a ring-like, perhaps clumpy dust distribution. We detect low-level star formation in the inner parts of the gas disk, and find traces of AGN activity through increased ionization rates.}

The ISM components are distributed in a nested structure of increasing spatial extent. The molecular gas resides closest to the galaxy's center, within the radius of the dust lane. Beyond that we find the ionised gas, with an interesting tail towards the South side of the galaxy and slightly blueshifted with respect to the main ionised disk. The position and relative velocity of FCC~167 to the cluster center indicate that the galaxy is moving through the cluster on its way to apocenter.  
\textit{Two very different scenarios emerge to explain this nested ISM structure. On the one hand, gas could be falling into the central potential of FCC~167. The ionised gas gradually settles in a ring where it can cool down and form molecules and stars in the shielded inner regions However, this scenario is unlikely to happen in a cluster environment. On the other hand, the current ISM could be a relic of a far greater reservoir that pre-dates the galaxy's interaction with the cluster environment. The gas and dust can gradually be stripped by ram pressure, exposing the dust to sputtering and ionizing the gas.}

We performed 3D radiative transfer simulations to fit the intrinsic structure of the dust and break the geometry-dust mix degeneracy that shapes the attenuation curve. \\
\textit{A model without small carbonaceous grains fits the optical attenuation much better than the Milky Way dust mix. Still, both models underestimate the FIR dust emission by $\sim67 \%$. We attribute the energy deficit to a clumpy dust distribution after ruling out a second, diffuse, dust component.} \\
Our interpretation is supported by the high line-to-continuum reddening ratio, and by efficient dust destruction timescales. Indeed, hot gas is present in the central regions of this galaxy. A fit to the X-ray spectrum yield gas properties that would destroy the dust grains by sputtering on the order of an orbital timescale ($\sim7$ Myr). \\
\textit{Efficient dust destruction explains why almost no small grains are present in the dust lane, and requires the remaining dust to be self-shielding if the ISM is to survive.}

Our study has revealed that we may be witnessing a very particular moment in the history of FCC~167, where dust and gas are quickly processed as they are exposed to a hostile environment. Our findings show how detailed analysis of individual systems can complement statistical studies of dust-lane ETGs. Together, these studies can pinpoint the origin of the ISM and place dust-lane ETGs on the general timeline of galaxy formation.

\begin{acknowledgements}
S.V. is supported by the UGent-BOF fund and the UHerts visitor fund. E.M.C. and L.M. acknowledge financial support from Padua University through grants DOR1699945/16, DOR1715817/17, DOR1885254/18, and BIRD164402/16. GvdV acknowledges funding from the European Research Council (ERC) under the European Union's Horizon 2020 research and innovation programme under grant agreement No 724857 (Consolidator Grant ArcheoDyn). \\ We thank Anthony Jones for several stimulating discussions.\\ Based on observations collected at the European Organisation for Astronomical Research in the Southern Hemisphere under ESO programme 296.B-5054(A). \\ This research has made use of data obtained from the Chandra Data Archive. \\ This paper makes use of the following ALMA data: ADS/JAO.ALMA\#2015.1.00497.S. ALMA is a partnership of ESO (representing its member states), NSF (USA) and NINS (Japan), together with NRC (Canada) and NSC and ASIAA (Taiwan) and KASI (Republic of Korea), in cooperation with the Republic of Chile. The Joint ALMA Observatory is operated by ESO, AUI/NRAO and NAOJ.

\end{acknowledgements}

\bibliographystyle{aa} 
\bibliography{allreferences}

\end{document}